\documentclass[12pt]{iopart}
\usepackage[T1]{fontenc}
\usepackage[utf8]{inputenc}
\usepackage{graphicx}
\usepackage{color}

\newcommand{\cblue}{\textcolor{black}}
\begin{document}

\title[Physically motivated projection of the ECG]{Physically motivated projection of the electrocardiogram - a feasibility study}

\author{Sebastian Wildowicz$^1$, Tomasz Gradowski$^1$, Paulina Figura$^2$, Igor Olczak$^1$, Judyta Sobiech$^1$, Teodor Buchner$^1$}
\address{$^1$Faculty of Physics, Warsaw University of Technology\\$^2$Children's Memorial Health Institute}
\ead{sebastian.wildowicz.dokt@pw.edu.pl}
\vspace{10pt}

% \linenumbers

\begin{abstract}
We present PhysECG: a physically motivated projection of the 12 lead electrocardiogram, supported by a deep learning model trained on 21,799 recordings from the PTB-XL database and discuss its feasibility. The method allows to evaluate the epicardial activity (inverse problem of ECG imaging) and, in particular, to distinguish left and right ventricular activity, with statistical spread
related to localization of the septum. The observed dyssynchrony resembles other experimental results. The foundations of the method are based on the molecular theory of biopotentials. The heart's activity in view of the method is decomposed into two processes: the passage of the electric activation wavefront and the response of cardiomyocytes. We introduce the idea of the electrode-resolved activity function, which represents the mass of the ventricle in Phase 0 of action potential within the lead field of each electrode. The computations are fast and robust, with excellent convergence. We present the quality metrics for the reconstruction based on the model on the testing set selected from the PTB database. In order to prove feasibility, we present and discuss two healthy controls: male and female, and two pathologies: right bundle branch block, and anterior myocardial infarction. The results obtained using PhysECG seem to be in accordance with the changes evoked by pathology, which has to be confirmed by subsequent clinical studies. The method is based on ECG, and does not require reconstruction of body geometry, which presents an affordable solution for low and middle-income countries where access to imaging is limited. 

\end{abstract}

\newpage
\section{Motivation}
\label{sec:motivation}
The 2023 Guidelines for the management of acute coronary syndromes \cite{Byrne2023} point at cardiovascular diseases (CVD) as the most common cause of mortality and morbidity in the world population, with an exceptionally high burden carried by low- and middle-income countries (LMIC). The diagnostic tool with a straightforward and robust measurement protocol is the electrocardiography (ECG), for which Willem Einthoven was duly awarded the Nobel Prize.
Unfortunately, the simplicity of the ECG is ostensible, as its interpretation still requires high qualifications, which are not always accessible, especially among the LMIC.

The advent of computer-assisted interpretation techniques to support the opinion of a clinician brings a major breakthrough and may significantly improve the quality of ECG diagnosis in real-life scenarios \cite{Kashou2023}.

Among computer-assisted techniques, two attract particular attention: machine learning (ML), which is quite pompously called artificial intelligence (AI), and ECG imaging (ECGI). It was only a matter of time before we saw the first applications of ML to the interpretation of ECG. Generally, clinicians tend to cool down their tempers, pointing at the need to thoroughly verify the otherwise promising results \cite{Ose2024, Niemczyk2024, Holmstrom2024, Kolk2024}. The last of these papers by Kolk et al. \cite{Kolk2024} is particularly interesting, as it points out the importance of a multimodal approach, including various imaging modalities. It is not surprising to find that such sophisticated techniques as late gadolinium enhancement (LGE) or cardiac magnetic resonance imaging (CMR) bring an additional angle of observation, which significantly improves the quality of sole ECG-based automatic diagnosis \cite{Kolk2024}.

The rising trend to use the multimodal approach in cardiological diagnosis was also observed by the authors of the guidelines, e.g., 2023 ESC Guidelines for the Management of Acute Coronary Syndromes (ACS) \cite{Byrne2023}. The interest of clinicians moves from the sole ECG-based diagnosis towards a multimodal one: in the ACS, the ultrasound is considered more important than exercise ECG (although both are still in class IIa~-~c.f. \cite[p. 3742]{Byrne2023}. Bacharova, {and the group of electrocardiology experts}, pointed at multimodal methods as the solution for clinical assessment of cardiac hypertrophy \cite{Bacharova2023}. Hypertrophic cardiomyopathy (HCM), dubbed "the great masquerader" \cite{Maron2018}, is estimated to be undiagnosed in 85\% of cases \cite{Martinez2024}. In cardiac amyloidosis, a recently identified root cause, the resting ECG presents a particularly low sensitivity of 28\% \cite{Cheng2012}. Diagnosis of non-ST-elevation myocardial infarction (NSTEMI) points at biochemical and blood pressure-related markers that substitute findings invisible in the ECG \cite{Collet2021}. Embarrassingly, the prevalence of sudden cardiac death (SCD) \cite{pmid29084731,pmid27029760} is still too high, which means, i.a., that we cannot effectively predict its occurrence. However, due to ML, the situation seems to gradually improve \cite{Holmstrom2024, Kolk2024}. It is daring to say that despite many years of rapid development and a helping hand from the ML algorithms, electrocardiological diagnosis is in crisis.

This is more than true for the LMIC. Unfortunately, virtually all imaging techniques require costly equipment, so the multimodal solutions are naturally biased toward high-income countries. We propose to reconsider the ECG for two reasons: 1) it remains the most popular diagnostic method in the world, particularly useful in LMIC due to the low cost of equipment, 2) we believe that the diagnostic capabilities of the ECG were not drained yet and introduction of new ML-based techniques may provide more valuable clinical data.

One example of an unsolved problem with the ECG is the inability to discern between left ventricular and right ventricular ectopy. Localizing the source of arrhythmia is crucial, as it determines how the ablation procedure is performed \cite{Anderson2019}. The functional state of cardiomyocytes, whether we analyze electrical stability or transcriptomics \cite{Frisk2024} seems to be a quantity of high spatial variability, which requires the methods that could provide precise localization. An interesting method of determination of QRS outline does not allow to differentiate the left and right ventricular source \cite{Sedova2023}, further methods of localization will be discussed later.

This is a perfect starting point to describe the next actor on this stage: the Electrocardiographic Imaging techniques (ECGI) \cite{Cluitmans2018}. These techniques present a physics-informed imaging modality, which aims to reconstruct cardiac potentials from the surface ECG \cite{Li2024}. Substantial domain knowledge was used in order to build a relatively low-expense method, which brings the power of imaging to cardiological diagnosis. Unfortunately, this comes at the cost of complex mathematics, which often leads to poor convergence and, in consequence, the low spatial resolution of cardiac sources. The ability to localize the source, based on sole ECG, still needs to be improved \cite {Bear2018} - this is currently the main drawback of a promising technique of electrocardiological imaging. One of the problems tackled by the ECGI is the mentioned problem of localization of ectopy, which makes it particularly worth considering. Body Surface Potential Mapping techniques are among the most promising \cite{Potyagaylo2019}, however, they are more problematic than 12 lead ECGs.
The definitive merit of ECGI, which also presents a particular drawback, is the ability to build personalized models - digital twins \cite{Li2024}. As it brings a substantial diagnostic yield, it also raises the bar concerning the total cost of diagnosis, which includes the magnetic resonance imaging (MRI) or computer tomography (CT) required to build a personalized model. This defines well-developed countries as the market for the idea of digital twins and other ideas of personalized medicine. Many societies are still underrepresented not only in clinical trials but also in access to modern therapeutic methods \cite{BibbinsDomingo}.

It has to be noted that a particular problem with reconstruction techniques may stem from the fact that the physical foundations of the widely used techniques are rarely discussed - c.f. \cite{Buchner2019,Buchner2023,Peczalski2024,Scharf}. There is a strong tendency to treat current foundations of the ECG: the cardiac dipole theory, the volume conductor theory, the bidomain model, and the idea of impressed current as ultimate \cite{bembook}. Here, we do not propose a headless shift in paradigm in the name of theoretical beauty \cite{Hossenfelder2019}. However, we do not hesitate to change the paradigm, wherever it is rational and brings a substantial scientific yield. As for the ECG imaging, we believe that widening the scope of methodology will improve the accuracy of reconstruction. We believe only in mathematical assumptions that are soundly physically motivated.

This sets a starting point for the current research. In our opinion, a promising solution is to combine the ideas of ECGI with the diagnostic power of AI on a physical basis, in order to improve the convergence of reconstruction methods, and their diagnostic potential. Currently, we limit ourselves only to the 12 lead ECG data in a standard electrode setup. This work aims to show that the AI-based implementation of the PhysECG algorithm for evaluation of cardiac activity wave from the surface electrocardiogram is a viable alternative for other ECG methods, which comes at particularly low operational costs, that allows its usage in the LMIC operational theater of cardiological diagnosis. The work merely presents this technique's feasibility on selected examples, as its full clinical development requires further study.

The structure of the paper is as follows. We describe the algorithm for ECG synthesis from cardiac sources (forward problem \cite{pmid26082525}), which is the easiest way to present the inverse problem \cite{Li2024, Cluitmans}. Then, we present the architecture of a deep learning model and the results obtained from its use. We summarize the paper by discussing the results and conclusions obtained.

\section{Theoretical background}
\label{sec:process}

The algorithm we present is grounded on the molecular theory of biopotentials (MTB), which is briefly outlined in \ref{app:Molecular}.
Among its most important consequences is the assumption that the potential that spreads over the passive tissue is unipolar, and the bipolar nature of the ECG comes from the fact that it represents a space gradient of unipolar potentials. This has been directly shown for the myocardium by multiple authors contributing to monophasic potential measurements \cite{Franz1999}. In consequence, the ECG measures the asymmetry of unipolar electrode potentials. Of course, it is widely acknowledged that the measurements in limb leads are bipolar. Nevertheless, it is tough to imagine that the potentials that spread over the body reveal only a tiny part of their signal power, known as the ECG. It is sufficient to firmly press one of the electrodes to the myocardium, turning it into a temporary reference point, to reveal the unipolar nature of the biopotential \cite{Franz1999}. Further discussion of this subject is presented in Discussion section.
As a consequence of the MTB, we can assume a direct relation between the epicardium and the electrodes, based on the transport of ions and spatiotemporal evolution of volume polarization vector. This type of relation is the core of a mathematical projection commonly used in forward problem \cite{Barr1977,BAMS2024}.

We propose to decompose the process of passage of the biopotential through the epicardium into two distinguishable physical processes: P1 and P2.

Process P1 is the spatiotemporal motion of the activation wave, understood as a set of loci that enter the active state (Phase 1 of action potential (AP) ) at a given time instant. This process may be described using isochrones of activity, similar to those obtained by Durrer \cite{Durrer} and in numerous experiments related to ECGI \cite{Li2024, Cluitmans2018}.

Process P2 is the response of an individual cardiomyocyte to the stimulation. It is naturally related to the action potential duration (APD) for each cardiomyocyte and therefore to the depth of the activation front. 

P1 and P2 may be decomposed only as the first approximation, as many relations between them are known. The heart rate (HR) affects the APD via APD/DI relation \cite{Orini2016}, the curvature of the wavefront affects the propagation velocity \cite{Fast1997}, and APD and its restitution, in turn, depend on the curvature of an activation front \cite{Hanson2009, Qu2000} and determine its stability \cite{Weiss1999}.

Also, the morphology of T end is directly related to the velocity of the activation wave \cite{Jabr2016}, as the ionic pool inside each cell obeys the charge continuity equation:

\begin{equation}
\nabla \cdot \mathbf{J} + \frac{\partial \rho}{\partial t} = 0
\end{equation}
where \( \mathbf{J} \) is the current density and \( \rho \) is the charge density.

Despite many interrelations between P1 and P2, it seems, that they can be neglected, at the lowest level of approximation. The relations mentioned above may always be plugged in at a later stage. We present the algorithm below, being fully aware of the extent to which the physical processes are simplified, but we tend to prove that the usefulness of this projection method justifies the approach.

\section{ECG Synthesis Algorithm}
\label{sec:algorithm}

For each locus at the epicardium, it is possible to determine, with some accuracy, a coupling matrix that evaluates its contribution to the biopotential of the electrode \cite{Barr1977,BAMS2024}. These techniques are used in cardiac function imaging, specifically in potential mapping methods. Therefore, the signal arriving at the electrode results from the activation of a specific volume of cardiomyocytes. Further, it can be assumed that the epicardial potential is just a superposition of AP of cardiomyocytes, which directly propagates to the skin as a unipolar potential (c.f. \cite{Franz1999}). In that case, we can apply a standard convolution operator, taking the cardiac AP as basis function $m(t)$, as detailed below. In the studies described in this paper, we imported to Myokit the ten Tussher model from CellML \cite{Cuellar} to generate basis functions \cite{Ten_Tusscher2004-ns}. However, this choice was arbitrary, as we have also used Luo Rudy \cite{LuoRudy} provided by Myokit \cite{Clerx2016}. 

\begin{figure}[ht]
\centering
\includegraphics[width=0.5\linewidth]{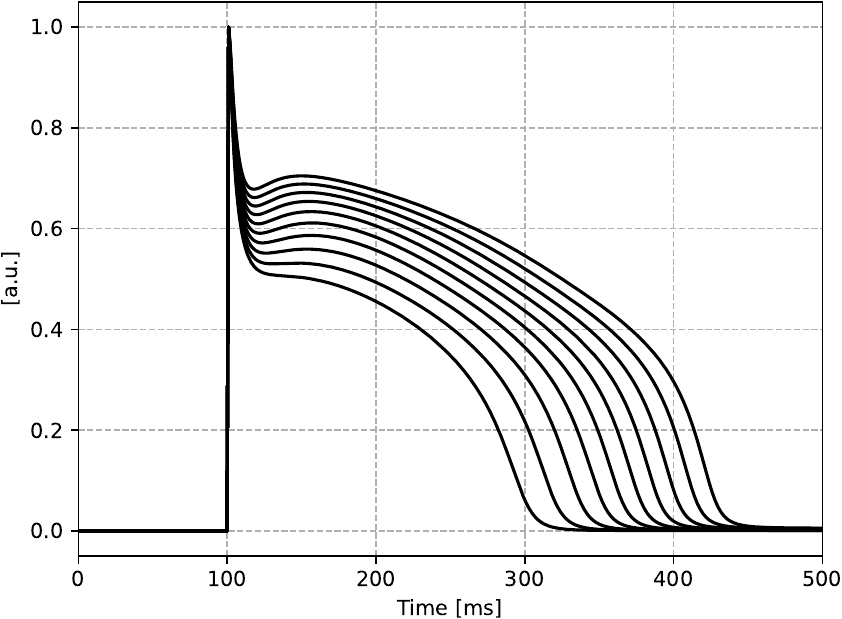}
\caption{Series of cardiomyocyte action potentials of Luo-Rudy model \cite{LuoRudy} generated using Myokit \cite{Clerx2016}. The series were obtained by multiplication of the parameter ib.gb=0.03921, by a factor of 1.5 to 2.5. All other parameters have default values.}
\label{fig:LR}
\end{figure}

The action potential of this model is presented in \cblue{Fig.} \ref{fig:LR}, for different parameter values to illustrate the effect of inotropic regulation on the action potential waveform. This regulation occurs naturally through the phosphorylation of channel proteins \cite{Jabr2016} and affects the electric and mechanical response of the cell. 

As a result of the above considerations, we assume that any unipolar potential $\Phi_i(t)$ (where $i$ is the electrode number) can be expressed using the convolution operator, and basis functions $m(t)$. The following equation mathematically describes the decomposition into processes P1 and P2:
\begin{equation}\label{eq:projection}
\Phi_i(t) = m(t) * k_i(t) = \int m(t-\tau) k_i(\tau) d\tau
\end{equation}

where:

\begin{itemize}
    \item $\Phi_i(t)$ represents the unipolar potential at electrode $i$,
    \item $m(t)$ is the base function representing cardiomyocyte action potential (P2),
    \item $k_i(t)$ is the mapping kernel for electrode $i$, further referred to as activity function (P1).
\end{itemize}

Convolution ($*$) in eq. \ref{eq:projection} is a mathematical operation that combines two functions to produce a third function, illustrating how the shape of one of them is modified by the other. Thus, one of the functions may be considered an impulse response of a filter through which the other function passes. This is exactly the case for the ECG: the activity function (P1) acts as a filter through which the cardiomyocyte potential (P2) passes to produce unipolar electrode potential. The unipolar potential $\Phi_i(t)$ is hard to observe with bipolar techniques since it is measured against an ideal zero reference, which is difficult to achieve in living organisms without special techniques, such as monophasic recordings \cite{Franz1999} or utilization of a nonstandard electrode setup \cite{Gargiulo}. However, it directly relates to electric unipolar signals of the myocardium \cite{Franz1999} and to optical mapping measurements that allow to visualize the myocardial potentials using the voltage sensitive dye \cite{Kappadan2023}.

The concept behind $k_i(\tau)$, which formally represents the kernel of the convolution operator, involves the local distribution of activation times for those epicardial segments that contribute to projection to the specific electrode. This function is proportional to the muscle mass activated by the activation front at any given time (P1). It is not an actual distribution since it is not normalized to unity. There is no exact equivalence between muscle mass and signal amplitude, only proportionality, allowing for inference about temporal changes. The values are not absolute, as many factors, including skin condition under the electrode influence the amplitude. 

It is important to note, that the decomposition defined by eq. \ref{eq:projection} applies to epicardial potentials and consequently to their propagation to the torso. We assume, that the epicardial potentials $\Phi(\vec{r},t)$ may be represented as a convolution of: a) global activation front $T(\vec{r},t)$, which comprises of all points at epicardium, that enter the Phase 0 of action potential at certain time instant and b) a cardiomyocyte response function $m(t)$. What we observe at electrode is the electrode-resolved function $k(t)$, which represents the total mass of the part of activation front $T(\vec{r},t)$, which passes through the lead field of the electrode. Concerning the relation of this decomposition to the boundary element methods, that allow to solve the forward problem, it is such, that they define a geometric relation between points on epicardium and a  points on the thorax, that is invariant in time and solve Laplace problem for a given quasistatic boundary condition, which comprises of epicardial potentials at a certain time instant \cite{Barr1977,BAMS2024}. The electrode potential is defined as surface average over a fragment of a torso boundary, over which the electrode is located. The geometric relation mentioned above defines an effective lead field for each electrode. At each time instant, at the electrode, the projected epicardial potential is integrated. In fact, we decompose this potential into the time-variable activation mass, within the lead field (P1), and cardiomyocyte potential (P2). A precise mathematical formulation of the method is being developed.

From a formal perspective, a cardiological lead, such as the simplest limb lead, is the difference between unipolar potentials, equivalent to the spatial gradient, representing the asymmetry in the source itself and the propagation process. The commonly overlooked bipolar nature of the ECG (even in precordial leads that are not really unipolar \cite{OKAMOTO1998291, Gargiulo}) significantly complicates the diagnosis based on ECG. The activation time distribution, which can be estimated using reconstruction techniques such as electrocardiographic imaging (ECGI), exhibits many intriguing features that constitute a rich research program, as shown below.

The unipolar potentials decomposed with eq. \ref{eq:projection} may be transformed into standard ECG leads: Einthoven, Goldberger, and Wilson, as detailed below.

\subsection{Definitions of Leads}

The standard Einthoven limb leads, fundamental for traditional ECG recording, are defined as the differences in potentials between specific electrodes placed on the limbs of the patient:

\begin{eqnarray}\label{eq:limb}
I &=& \Phi_{LA} - \Phi_{RA} \label{eq:eq2} \\
II &=& \Phi_{LL} - \Phi_{RA} \label{eq:eq3} \\
III &=& \Phi_{LL} - \Phi_{LA} \label{eq:eq4}
\end{eqnarray}

where $\Phi_{LA}$, $\Phi_{RA}$, and $\Phi_{LL}$ represent the unipolar potentials of the electrode on the left arm, right arm, and left leg, respectively. Using projection of eq. \ref{eq:projection}, each of these leads has its corresponding activity function $k_i(t)$.

%\subsection{Determination of Goldberger Leads}

Goldberger augmented limb leads are used to obtain a more detailed view of cardiac activity. These leads are calculated as the differences between one electrode and the average of the other two limb electrodes:

\begin{eqnarray}\label{eq:goldberger}
aVR &=& \Phi_{RA} - \frac{\Phi_{LL} + \Phi_{LA}}{2} \label{eq:eq5} \\
aVL &=& \Phi_{LA} - \frac{\Phi_{RA} + \Phi_{LL}}{2} \label{eq:eq6} \\
aVF &=& \Phi_{LL} - \frac{\Phi_{RA} + \Phi_{LA}}{2} \label{eq:eq7}
\end{eqnarray}

where $aVR$, $aVL$, and $aVF$ represent the augmented leads for the right arm, left arm, and left leg, respectively.

%\subsection{Determination of Wilson Precordial Leads}

To generate the precordial leads $V_1$ to $V_6$ according to the Wilson system, a reference called Wilson Central Terminal (WCT) is used \cite{WILSON1934447}, which was at the time believed to make it possible to observe unipolar leads. Now we know that this is no longer true \cite{OKAMOTO1998291, Gargiulo}. WCT is the average of the potentials from the three limb electrodes:

\begin{equation}\label{eq:wct}
\Phi_{WCT} = \frac{\Phi_{LA} + \Phi_{RA} + \Phi_{LL}}{3} 
\end{equation}

Each precordial lead $V_i$ is then determined as the difference in potential between the precordial electrode and WCT:

\begin{equation}\label{eq:precordial}
V_i = \Phi_{V_i} - \Phi_{WCT},\,i=1,...,6 
\end{equation}

where $\Phi_{V_i}$ is the unipolar potential at precordial electrode $V_i$.

With the help of equations \ref{eq:eq2} to \ref{eq:precordial}, the projection equation \ref{eq:projection} can be used to synthesize the ECG from the given activity functions: one per each electrode. The next chapter describes the method of determination of activity functions from the ECG.

\section{Deep learning model for ECG analysis}
\label{sec:model}

The next step was to develop a deep learning model, which could enable us to generate a set of $k(t)$ from any previously recorded 12-lead ECG. The unsupervised learning scenario may be used, as the role of the model is to mimic the real ECG signal best. This gives a natural error function, a mean square error (MSE) between the ground truth and the reconstructed ECG.

\begin{figure}
\centering
\includegraphics[width=0.25\textwidth]{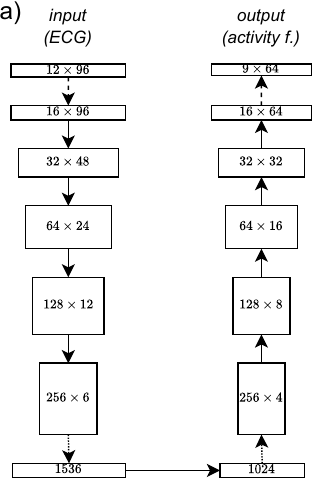}
\hspace{2cm}
\includegraphics[width=0.45\textwidth]{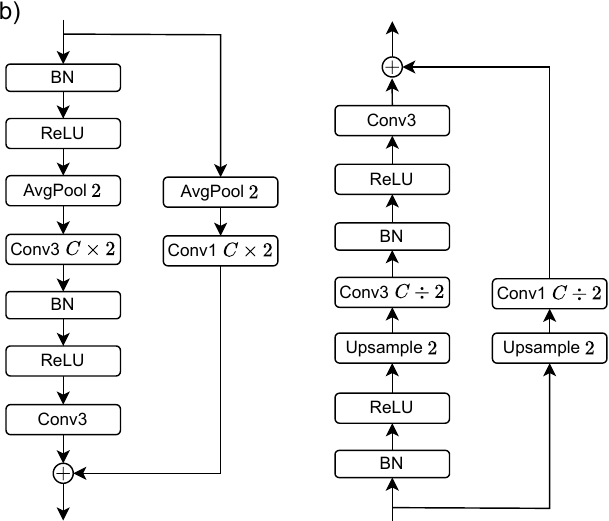}

\caption{Deep learning model architecture. a) signal flow in the model from the input (ECG signal) to output (activity functions) through the encoder (left branch) and decoder (right branch) subnetwork. Each rectangle represents the tensor of channels $\times$ length of the feature maps or flattened vector in the bottleneck. Solid vertical arrows represent residual blocks, dashed arrows represent projection blocks (convolutional layer with a kernel of size 7 followed by batch normalization layer and ReLU function in the encoder and single convolutional layer with a kernel of size 1 in the decoder), and dotted horizontal arrows represent the bottleneck transformation (fully connected layer + batch normalization + ReLU). b) structure of the residual blocks: downsampling block (used in the encoder subnetwork) on the left and upsampling block (used in the decoder subnetwork) on the right. Abbreviations and symbols: BN - batch normalization, ReLU - rectified linear unit, Conv$n$ - convolutional layer with kernel length of $n$, $C \times 2$ and $C \div 2$ - doubling or halving respectively the number of features, AvgPool 2 - average pooling layer with kernel of 2, Upsample 2 - nearest neighbor upscaling interpolation by a factor of 2, $\oplus$ - element-wise addition.}
\label{fig:model}
\end{figure}

The deep learning model for PhysECG analysis utilizes a compact, efficient convolutional encoder-decoder architecture designed to extract activity functions from 12-lead ECG signals (Fig.~\ref{fig:model}). The model processes 96-sample ECG segments (384 ms at 250 Hz) and outputs nine activity functions (one per electrode) of 64 samples each, enabling an efficient and scalable implementation with a total of 2.26 million trainable parameters.
The model consists of three main components:
\begin{itemize}
    \item Encoder: Downscales input signals while doubling feature map dimensions through convolutional layers, batch normalization, and average pooling.
    \item Bottleneck: Reduces dimensionality to capture critical features using a fully connected layer with nonlinearity (ReLU).
    \item Decoder: Mirrors the encoder with upsampling and deconvolutional layers to reconstruct activity functions.
\end{itemize}
This design follows principles of residual learning, employing skip connections for better convergence and stability while maintaining a small memory footprint. The entire model is implemented in PyTorch \cite{pytorch} and trained on the PTB-XL \cite{ptbxl} database with unsupervised learning, using mean squared error (MSE) between reconstructed and ground-truth signals as the loss function.

The model was trained on the PTB-XL database, which consists of 21,799 12-lead ECG recordings of 10 seconds obtained from 18,869 patients. The signals were resampled from the original sampling rate of 500 Hz to 250 Hz. The signal preprocessing and detection of the R peaks were performed using the Neurokit package \cite{neurokit}. The training set consists of 253,752 segments of 96 samples (384 ms), each containing one QRS complex with an R peak at sample 20.

The model outputs nine activity functions $k$, one per electrode, which are then used to reconstruct the ECG signal, following the algorithm described in Sec. \ref{sec:algorithm}. First, we calculate the unipolar potentials $\Phi_i(t)$ for each electrode $i$ by convolving the activity functions with the myocyte action potential $m$, using the discrete convolution operator.
\begin{equation}
    \Phi_{electrode, t} = \sum\limits_{i=-31}^{31} k_{electrode, i} \cdot m_{t-i}
\label{eq:discrete_conv}
\end{equation}
We use zero-padding on the myocyte action potential to avoid the edge effects
and drop the last sample of the activity functions for the even length of the kernel. The unipolar potentials are then combined to form the bipolar leads according to the standard ECG definitions (eqs. \ref{eq:limb} to \ref{eq:precordial}). 
The loss function $L$ used in the training process is given by the equation \ref{eq:loss}, where $X_{lead, t}$ and $Y_{lead, t}$ are the elements of the matrices representing the input signal and the signal reconstructed from the obtained activity functions, respectively. 
\begin{equation}
L = \frac{1}{12(offset-onset)} \sum\limits_{lead=1}^{12} \sum\limits_{t=onset}^{offset} \left( X_{lead, t}-Y_{lead, t} \right)^2
\label{eq:loss}
\end{equation}
The sum is taken over all leads and samples in the QRS complex. We were interested only in reconstructing the QRS complex, leaving the repolarization analysis to future research. This goal was obtained by setting the limit of error function at samples from 0 to 40 ($onset$ and $offset$ values in eq. \ref{eq:loss}), which corresponds to 160 ms at 250 Hz sampling, centered on the R position, as detected by the arrhythmia analysis. The R position from lead I was used regardless of the QRS morphology and R position in other leads.

It has to be noted that due to the tiny size of the feature set, the learning process was very fast. Training of the model on RTX4070 GPU with 12GB of VRAM on the whole training set took approx. 6 hours. Notably, the learning was unsupervised, as the accuracy of reconstruction provided a natural loss function in this regard.

\section{Results}
\label{sec:results}

Having obtained the trained model, we started the testing phase with segments of ECG signals that were not used for training. The testing set was obtained from the PTB database \cite{ptb} (which does not overlap with the PTB-XL database used for training) and consisted of 74,088 QRS segments from 549 recordings from 290 subjects. Additionally, a subset of 10,681 segments from 80 recordings of 52 healthy patients was used to assess the model's performance on healthy subjects. The testing set was preprocessed in the same way as the training set, described in Sec. \ref{sec:model}, with the only difference being the downsampling of the signals from 1000 Hz to 250 Hz.

Concerning the general results of the testing phase of the trained model, we have obtained the mean square error (MSE) of the representation on the level of $0.003$ per sample ($0.002$ for healthy control group), which corresponds to an accuracy of $0.8 \%$ of total power ($0.5 \%$ for healthy control group). The mean Pearson correlation coefficient between the ground truth and the reconstruction was $0.9974$ ($0.9985$ for healthy control group). The outliers, in which the largest MSE was observed, were related to the low amplitude of recording in the V4 electrode (subject \#180, recording 0545), most probably due to poor electrode contact. Hence, the fault was on the side of the ground truth and not on the side of the reconstruction.

In order to provide a preview of the method's capabilities, for \cblue{this} feasibility study, we present the results for four typical patients who represent three different medical cases. They were selected arbitrarily from the test set.

\begin{figure}[ht]
\centering
\includegraphics[width=0.7\textwidth]{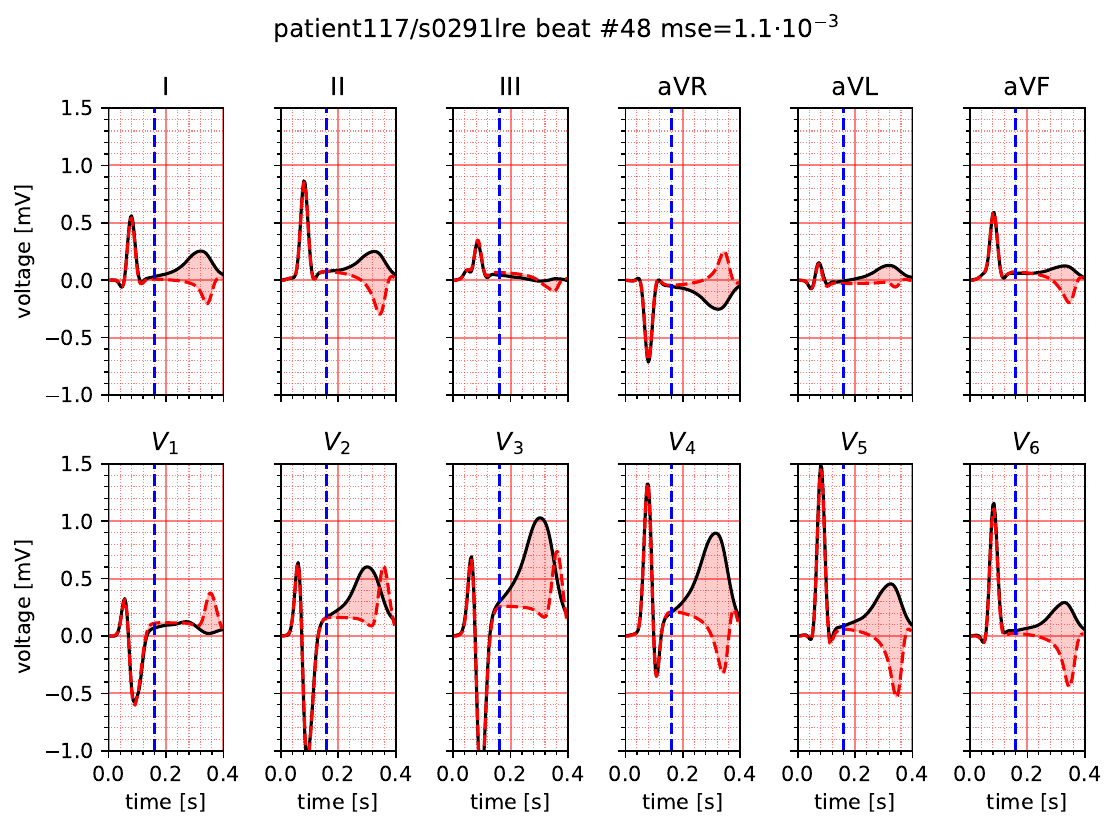}
\caption{Patient \#117 male, age 37, healthy control. Sinus beat of typical morphology in 12 lead projection and its reconstruction. Ground truth is in black, and reconstruction is in red. The area with red shading corresponds to the difference between the ground truth and the reconstruction. Note that the range limit in which the MSE for the loss function was calculated was at sample 40, at 80 ms past the R peak, i.e., immediately after the end of the QRS complex (blue dashed line). The differences in the ST region do not enter the reconstruction procedure.}\label{fig:117_ECG}.
\end{figure}

\begin{figure}[ht]
\centering
\includegraphics[width=\textwidth]{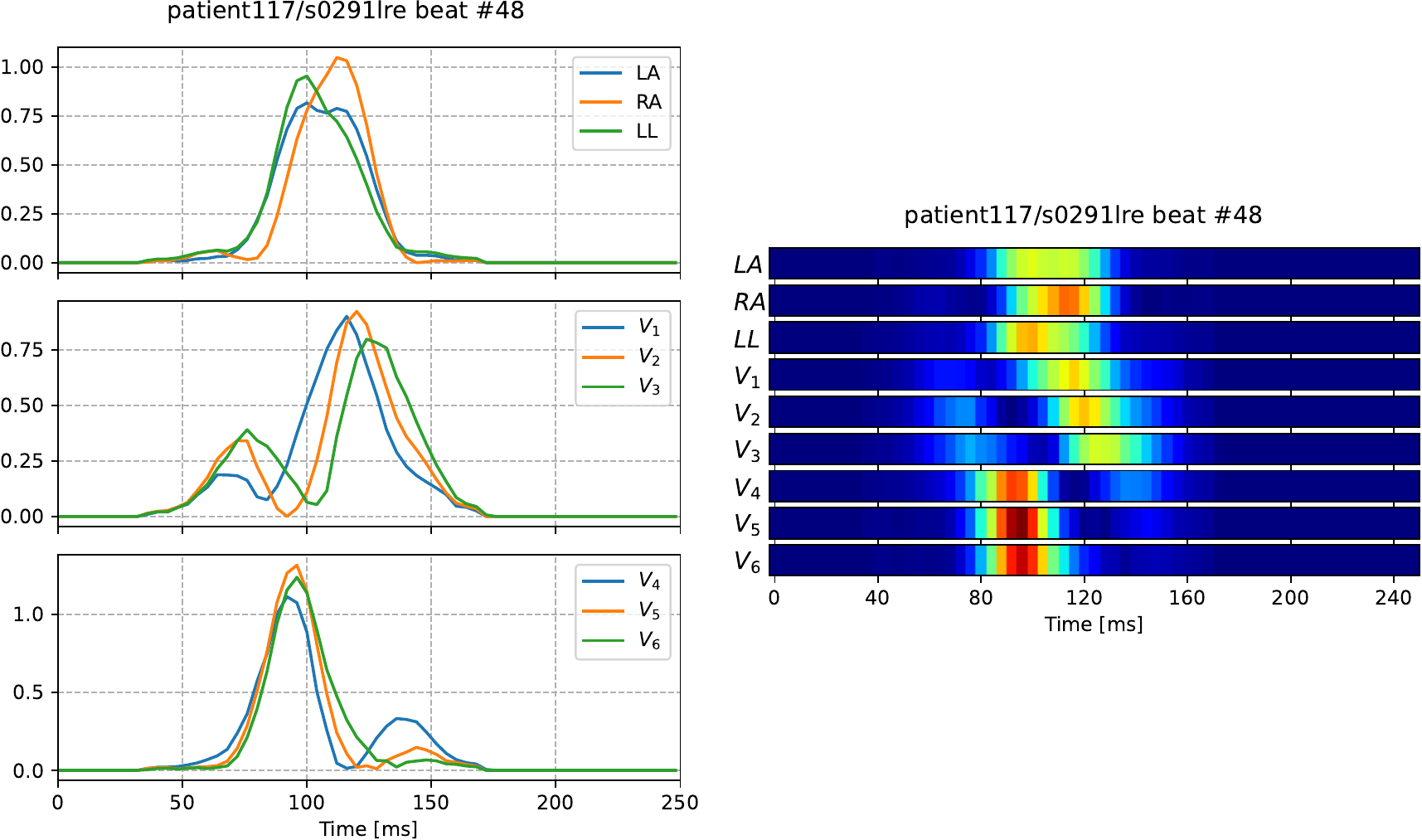}
\caption{Patient \#117 activity per electrode for a reconstructed beat as a function of time. The left maximum in precordial beats was identified as right ventricular activity, and the right maximum was identified as left ventricular activity. Ribbon projection on the right.}\label{fig:117_act}
\end{figure}

\begin{figure}[ht]
\centering
\includegraphics[width=0.7\textwidth]{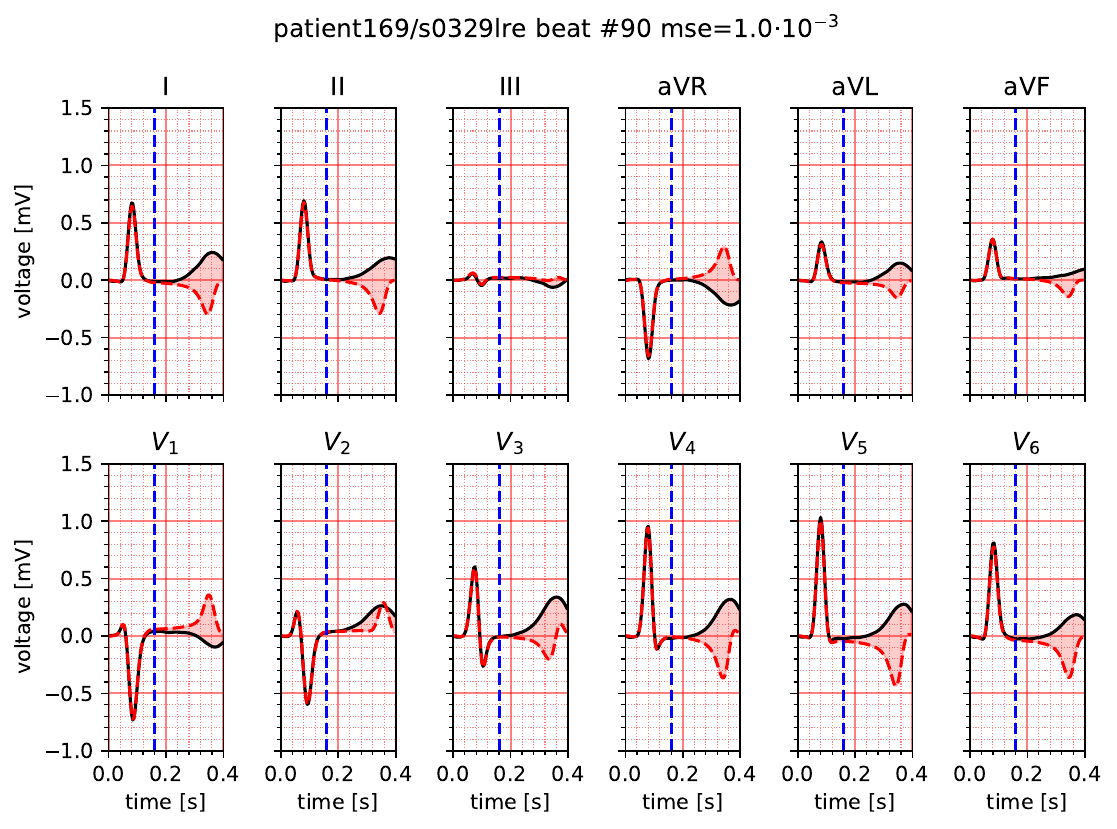}
\caption{Patient \#169 female, age 54, healthy control. Sinus beat of typical morphology in 12 lead projection and its reconstruction. Ground truth is in black, and reconstruction is in red.}\label{fig:169_ECG}
\end{figure}

\begin{figure}[ht]
\centering
\includegraphics[width=\textwidth]{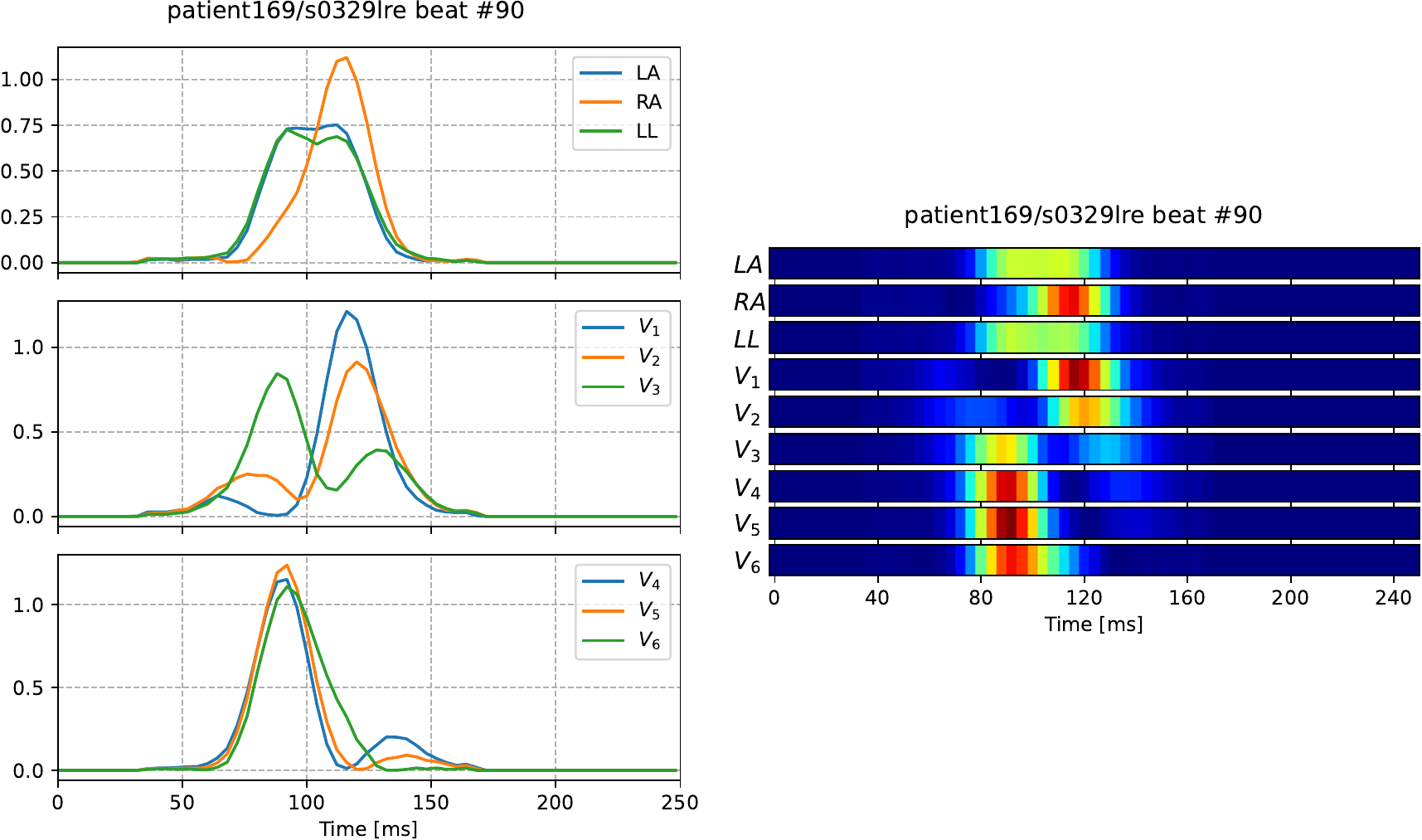}
\caption{Patient \#169 activity per electrode for a reconstructed beat as a function of time. Example of a recording with V3 dominated by left ventricle activity - left peak higher than the right peak. Ribbon projection on the right.}\label{fig:169_act}
\end{figure}

\begin{figure}[ht]
\centering
\includegraphics[width=0.7\textwidth]{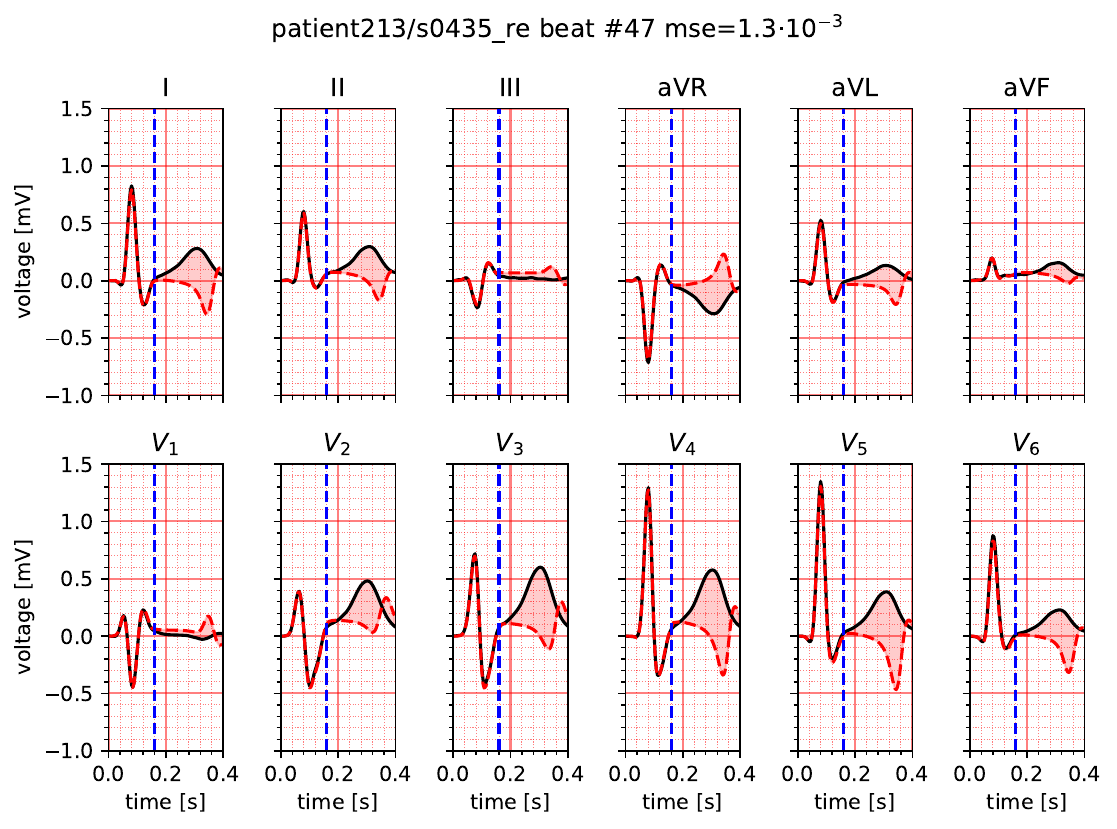}
\caption{Patient \#213 male, age 58, right bundle branch block symptoms. Sinus beat of typical morphology in 12 lead projection and its reconstruction. Ground truth is in black, and reconstruction is in red.}\label{fig:213_ECG}
\end{figure}

\begin{figure}[ht]
\centering
\includegraphics[width=\textwidth]{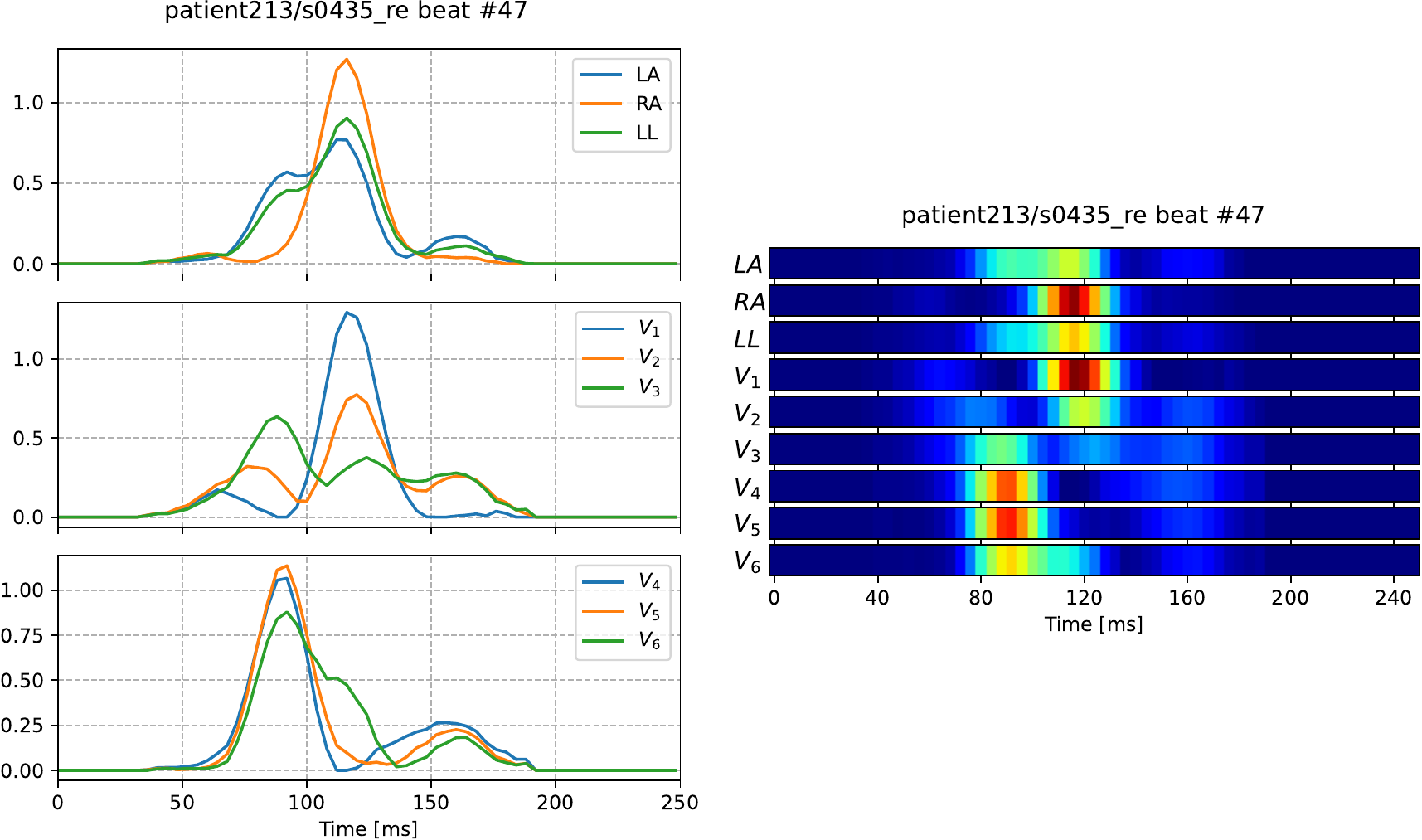}
\caption{Patient \#213 activity per electrode for a reconstructed beat as a function of time. Ribbon projection on the right.}\label{fig:213_act}
\end{figure}

\begin{figure}[ht]
\centering
\includegraphics[width=0.7\textwidth]{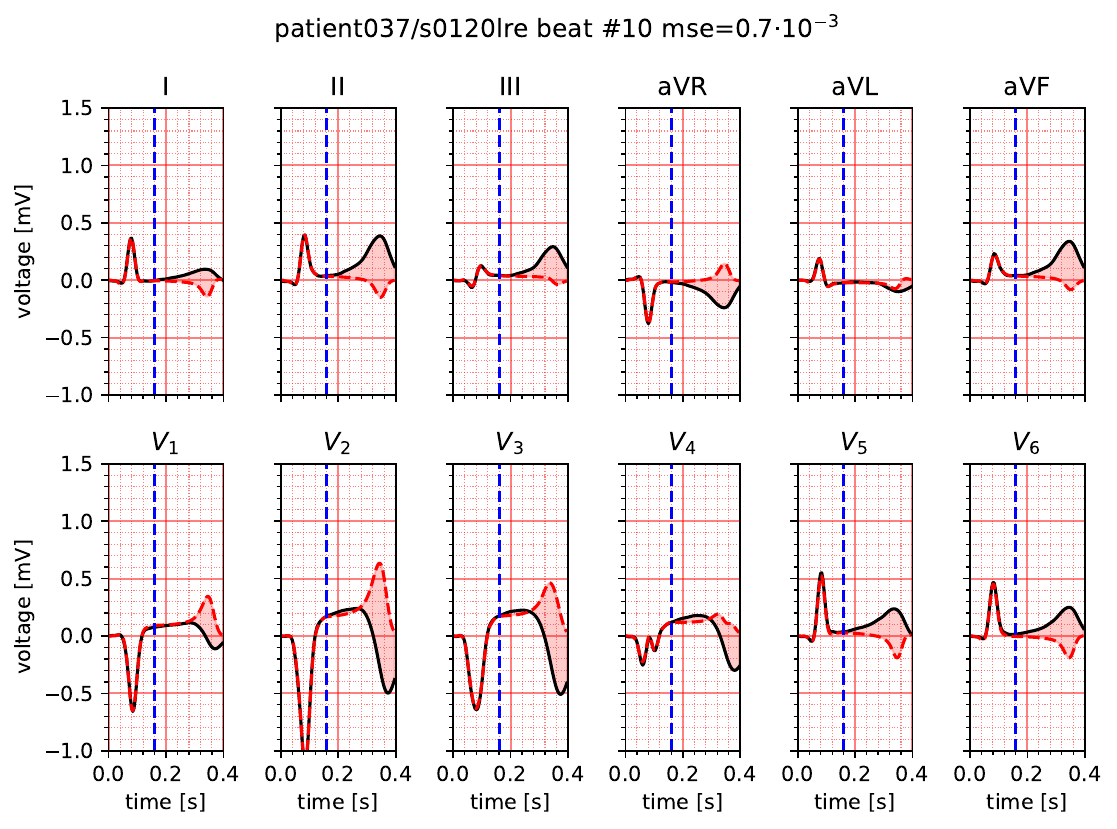}
\caption{Patient \#037 male, age 50, Anterior myocardial infarction, day 3 from the incident, symptoms of Pardee wave \cite{PARDEE} in V1-V4, medicated. 
 Sinus beat of typical morphology in 12 lead projection and its reconstruction. Ground truth is in black, and reconstruction is in red.}\label{fig:037_ECG}
\end{figure}

\begin{figure}[ht]
\centering
\includegraphics[width=\textwidth]{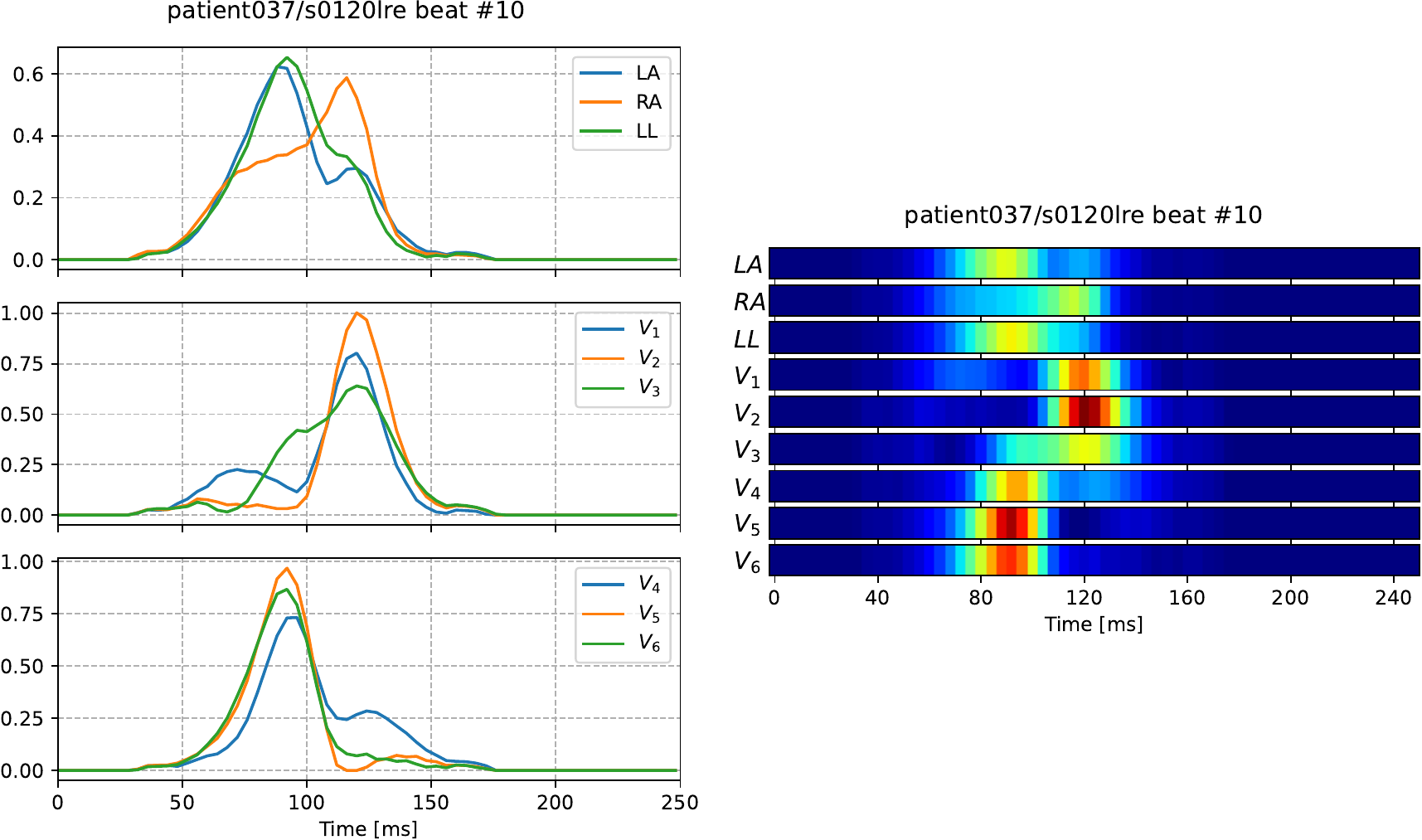}
\caption{Patient \#037 activity per electrode for a reconstructed beat as a function of time. Ribbon projection on the right. Note, that the vertical scale in top row is reduced, with respect to \cblue{Figs.} \ref{fig:117_act},\ref{fig:169_act} and \ref{fig:213_act}}\label{fig:037_act}
\end{figure}

We have selected elected results contain two controls: healthy male -  \cblue{Figs.} \ref{fig:117_ECG} and \ref{fig:117_act}, healthy female  - \cblue{Figs.} \ref{fig:169_ECG} and \ref{fig:169_act}, one example of Right Bundle Branch Block  - \cblue{Figs.} \ref{fig:213_ECG} and \ref{fig:213_act} and one example of Anterior Myocardial Infarction  - \cblue{Figs.} \ref{fig:037_ECG} and \ref{fig:037_act}. For each case, we present three figures: the reconstruction of ECG and two projections of activity function.

\begin{figure}[ht]
\centering
\includegraphics[width=0.6\textwidth]{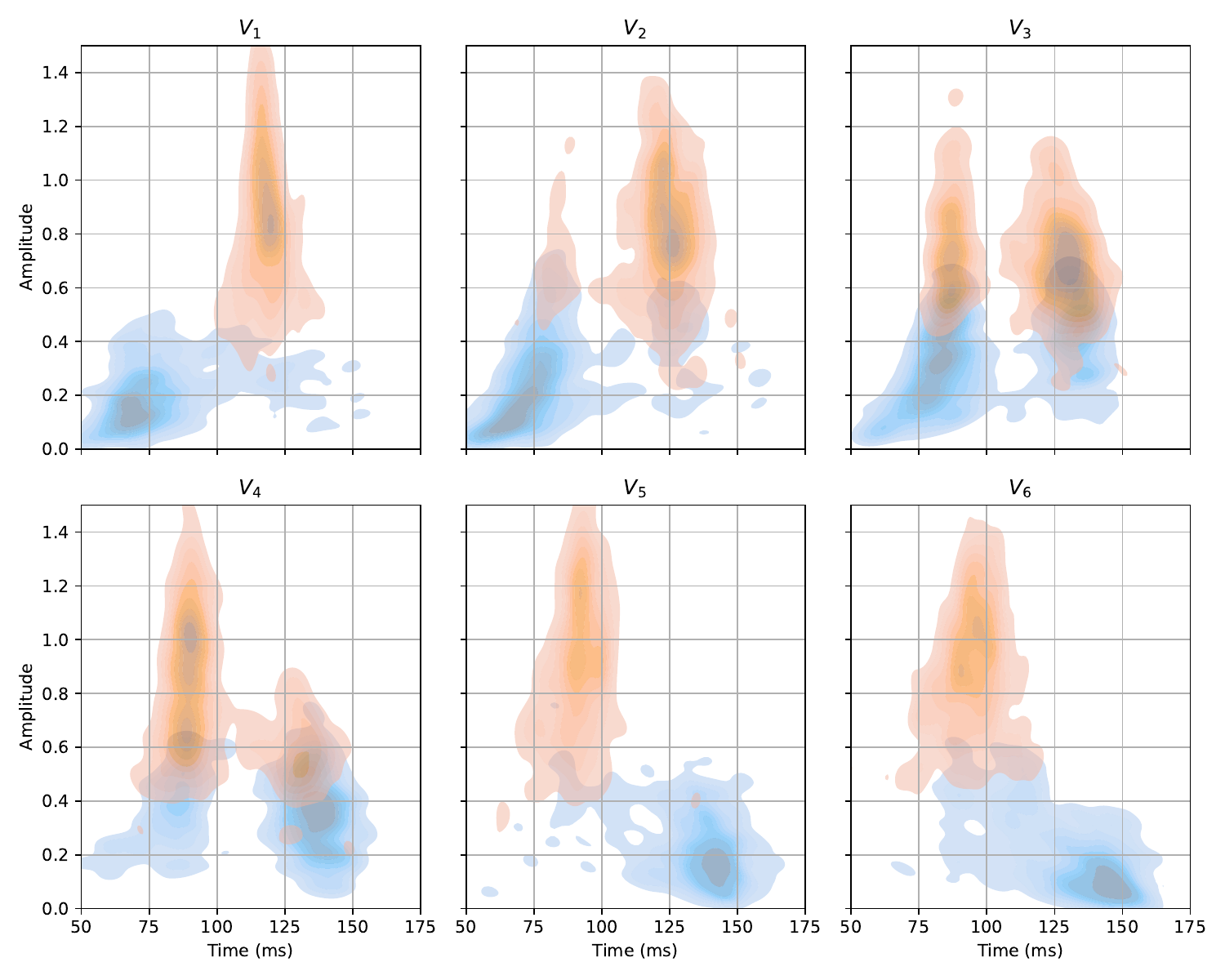}
\caption{Position and height of the first and second peak in precordial electrodes in position-amplitude projection for the testing set from the PTB database. A dominant peak in orange and a second peak in blue. Note that the position of the dominant peak in V3 is equally distributed between the first and the second position. Details in text.}
\label{fig:ptb_kde}
\end{figure}

Another general result we have obtained is the statistical analysis of the activity function parameters in the precordial electrodes for the whole testing set (52 healthy controls). We have calculated a fit using a sum of Gauss distributions: one of them would capture the dominant peak while the second one would capture the second peak. Each recording is represented as a dot in such a projection, in which the position of each maximum in time is on the horizontal axis, and its amplitude is on the vertical axis. The same projection is repeated for all precordial leads for all recordings from the testing group. The cloud of points was smoothed using Gaussian kernel density estimation from SciPy library \cite{2020SciPy}, and the result is shown in \cblue{Fig.} \ref{fig:ptb_kde}. It can be seen that the effect of the progression of the maxima between the electrodes appears in the whole testing group. This result is discussed in detail in the discussion section. The input data consisted of all beats of all recordings, which gives the data high statistical confidence.

\section{Discussion}
\label{sec:Discussion}

The discussion section is divided into four subsections that refer to the general statements (\ref{subsec:general}), and then to the examples of medical cases: healthy controls (\ref{subsec:healthy}), Right Bundle Branch Block (\ref{subsec:rbbb}) and Myocardial Infarction (\ref{subsec:mi}).

\subsection{General remarks}\label{subsec:general}

Our most significant doubts concerning the results that we have obtained were related to the observed split between ventricles (c.f. \cblue{Figs.} \ref{fig:117_act},\ref{fig:169_act},\ref{fig:213_act},\ref{fig:037_act}), which would mean that the wave of activity passes twice within the sensitivity field of a septal electrode, i.e., the one, the position of which is located over the point, where the septum has the closest distance (in an electric sense) to the epicardium. The projection we propose is based on a novel theoretical basis, so there were no other results to refer to.
However, one by one, our objections were mitigated. First, we found that the shape of unipolar potentials, which we postulate, was actually observed as monophasic action potentials \cite{Franz1999}; we also consulted the physiologists who still used them in 1970-ties, and referred to the lack of stability but not lack of usefulness (A. Beręsewicz, \textit{priv comm.}). Another confirmation of the shape, which spreads through the tissue, came from optical mapping techniques \cite{Kappadan2023}. Next, we have confirmed the dyssynchrony between the right and the left activation time \cite{Waddingham2022, Elliott2021}, which is often studied in the context of cardiac resynchronization therapy - c.f. \cite{Verstappen2024}.

Finally, we have found the series of papers by the Pavel Jurak group, devoted to a new measurement technique, the Ultra-High-Frequency ECG, which supports our result that the activity in precordial leads may be observed separately for each of ventricles \cite{Nguyn2024, Jurak2017}. The previous method of determination of the QRS envelope, developed by the same group, also resembles our results; however, it does not allow us to differentiate the left and the right ventricle \cite{Sedova2023}.
Separation of ventricular activity on the basis of standard 12-lead ECG seems to have a high diagnostic potential, which has to be confirmed by a study involving heart geometry. For the time being, we were only interested in reconstructing the QRS complex. Further development towards a clinically significant problem of cardiac repolarization \cite{Isaksen2021, KrijgerJurez2023} may include the approach proposed by Xue et al. \cite{Xue2010}. 

\subsection{Analysis of healthy controls}\label{subsec:healthy}

We selected the examples of healthy controls in a way that shows significant discrepancies related to activity in the V3 electrode. In \cblue{Fig.} \ref{fig:117_act}, it captures more activity of the right ventricle, whereas in \cblue{Fig.} \ref{fig:169_act}, the activity is dominated by the left ventricle.

Note that this projection clearly shows the anatomy-related variants of normal activity, depending on the rotation of the heart around the longitudinal axis of the body. We did not confront this finding with the MRI-derived anatomy. However, one may expect that this electric phenomenon is related to the position of the septum with respect to the axis of V3 on the transverse plane. The observed discrepancy will affect all the ECG-based measurements, which, in turn, will introduce variability that lowers statistical measures, such as sensitivity and specificity, reducing the value of ECG for cardiological diagnosis.

The analysis of the dependence of healthy controls ECG on the anatomy is an example of a straightforward interpretation, which is evident in the projection method we propose.

Note that the limb leads are imprecise, as they show the activity, which is averaged over the large mass of ventricles and potentially distorted by passage through the massive structures, such as the abdomen. 
%These examples show the clear advantage of precordial leads over limb leads. The Goldberger leads deliver only a better method to visualize the differences, but there is no new information compared to the limb leads. These remarks also present a clear educational value.

The presented results are typical of healthy controls. For instance, we have observed a reduction of activity function in V1 and V2 in females, which we putatively interpret as the result of anatomical differences between both sexes.
This shows that despite their direct interpretation, activity functions are subject to all the factors affecting electric measurements.

Concerning the progression of the maxima of activity functions $k_i(t)$ calculated throughout the testing group, that was reported in \cblue{Fig.} \ref{fig:ptb_kde}, it can be seen that this tendency has statistical power. It clearly follows, or rather precedes the known fact, that the polarity of the QRS progresses from negative to positive between V1 and V6. As the peak of left ventricle activation is delayed with respect to the right ventricle, a clear progression of the QRS polarity follows. Thorough verification of this fact would require a direct comparison to ventricular isochrones, one of the directions considered for future research.

\subsection{Analysis of Right Bundle Branch Block}\label{subsec:rbbb}

The analysis of the Right Bundle Branch Block case, presented in \cblue{Figs.} \ref{fig:213_ECG}, \ref{fig:213_act} is probably the most interesting due to the vibrant structure of activity functions, best visible in \cblue{Fig.} \ref{fig:213_act}. It is well known that the total duration of the QRS complex is wider in all BBB cases and exceeds 120 ms, which is the limit of physiological QRS duration \cite{Ikeda2021, Calle2022}. In the language of physical projection, it means that the duration of the activity has to be longer, and the activity function has to be lower, as the total percentage of the activated ventricle is still 100\%, and each cell is activated only once. From this, it follows that the activity function has to be low and broad, as the area under the curve has to be exactly the same as for healthy controls, as the sensitivity field of the electrode does not change due to the pathology of intracardiac conductance. Hence, the observed mass is always the same, but its distribution differs - c.f. \cblue{Figs.} \ref{fig:117_act} and \ref{fig:213_act}.
Also, the geometry of the passing wave, particularly its curvature, may only affect the ascending slope of the activity function but not the AUC. As the limb leads are considered, if we compare the RBBB to the healthy case, the RA is more delayed with respect to the LA and the LL - c.f. the ribbon projection in \cblue{Figs.} \ref{fig:213_act} and \ref{fig:117_act}, which makes perfect physiological sense.
The activity in left ventricular precordial leads resembles the healthy case - c.f. \cblue{Figs.} \ref{fig:117_act},\ref{fig:169_act}, as the physiological pathway functions properly - the left branch of the His Purkinje system. In comparison, the activity in right ventricular precordial leads has a lower amplitude and is clearly fragmented, and the fragments are separated in time. This is an electric manifestation of the altered geometry of the right ventricular activation wave, which was initiated from the left ventricle. As the velocity of conductance across the muscle is much lower than the one along the muscle due to anisotropy of the electric resistance tensor, there is a substantial delay between the separated waves. Further analysis of this case would require a direct comparison to isochrones from the electrophysiological (EP) study of the RBBB case, together with the ECG-based analysis of the physical projection, which is another direction of future research. Note also that the discrepancy in activation time between the neighboring lines of cardiomyocytes (that reside within the sensitivity field of a single electrode) is a factor that potentiates the progression of hypertrophy due to excessive shear stress within the muscle. In consequence, in prolonged BBB cases, indeed, a development of ventricular hypertrophy is observed \cite{Kloosterman2024}. To our knowledge, physical projection is the only one that would directly show the fragmentation of neighboring muscle volumes from the surface ECG. Of course provided, the results are validated by comparison with other imaging modalities, in particular the EP study providing true ventricular isochrones.

\subsection{Analysis of Myocardial Infarction}\label{subsec:mi}

The analysis of the MI case presented in \cblue{Figs.} \ref{fig:037_ECG} and \ref{fig:037_act} shows a reduction of LA and LL activity - c.f. \cblue{Fig.} \ref{fig:037_act} (note a difference in vertical scales explained in \cblue{figure} caption). These electrodes correspond to the anatomical location of the MI, which is located in the arterial wall of the left ventricle. Note that the activity of the right ventricle is unaffected, which also corresponds to the physiological reality. The physical projection clearly isolates the pathology source and shows its lateralization. As this is a preliminary study, we have not developed this observation towards full clinical confirmation with high statistical power, but it is an interesting action for future studies.

One may expect that the precordial leads better express the symptoms of the infarction, and it is well visible in \cblue{Fig.} \ref{fig:037_act}. The electrode which shows the largest changes is V3. The activity in V3 is broad, with reduced amplitude. Note that the shape of the activity function is affected not only by the reduction of the total mass of healthy responsive cardiomyocytes but also by alterations of the geometry of the activation wave, which does not pass through the infarction region (at least in case of STEMI infarction, and such is the case, which is clearly visible in \cblue{Fig.} \ref{fig:037_ECG}). Note that the right ventricle activity is also unaffected in precordial leads.

This example shows that, to a certain extent, the changes in the anterior wall also affect the leads, which are not directly related to the localization of the infarction region. This well corresponds with the results of many studies on statistical independence between ECG leads \cite{Pipberger1961-fi}. Further confirmation of this relation requires the analysis of infarction ECG for many cases with confirmed localization of the infarction site, which is one of the future research directions.

\section{Conclusion}
\label{sec:Conclusion}

We have presented the PhysECG algorithm - physically motivated ECG projection with help of the deep learning model trained on the PTB-XL database. It allows the decomposition of the heart's activity into two disjoint processes: the passage of the electric activation wavefront (P1) and the response of cardiomyocytes (P2). Process P2 is plugged into the model using ten Tussher model of ventricular cardiomyocyte response. Process P1 is evaluated through its activity function, which represents the mass of the ventricle in Phase 0 of action potential within the lead field of each electrode. We show a putative but hopefully convincing interpretation of the results of the model: the activity functions resolved per electrode that allows us to observe the moments of the passage of the ventricular activity wave in the lead field of each electrode separately. Using a standard electrode setup, we demonstrate the dyssynchrony between the right and the left ventricle. We also show the effects of population distribution of the anatomical position of the septum with regard to the direction of the V3 electrode on 52 healthy controls. We present selected results for male and female healthy controls: right bundle branch block, and anterior myocardial infarction and demonstrate that the results obtained using PhysECG are in accordance with the functional changes evoked by pathology. The algorithm may be used to reconstruct the activity on any 12 lead or Holter recordings. The method offers limited power of ECG imaging based entirely on sole ECG, which presents an affordable solution accessible for low and middle-income countries where access to resource-demanding imaging modalities is limited.

\section{Acknowledgements}
\label{sec:Acknowledgements}

A patent was applied for the PhysECG algorithm in Poland and abroad. R. Baranowski, M. Jastrzębski, J. Baran, J. Kanters,  J. Isaksen and A. Beręsewicz are thanked for valuable discussions. A. Staniszewska and M. Zajdel are thanked for their inspiration. P. Kuklik is thanked for his encouragement.

\section{Literature}
\label{sec:Literature}

\bibliography{references}

\begin{thebibliography}{10}
\expandafter\ifx\csname url\endcsname\relax
  \def\url#1{\texttt{#1}}\fi
\expandafter\ifx\csname urlprefix\endcsname\relax\def\urlprefix{URL }\fi
\expandafter\ifx\csname href\endcsname\relax
  \def\href#1#2{#2} \def\path#1{#1}\fi

\bibitem{Byrne2023}
R.~A. Byrne, X.~Rossello, J.~J. Coughlan, E.~Barbato~et al., 2023 {ESC} guidelines for the management of acute coronary syndromes, European Heart Journal 44~(38) (2023) 3720–3826.
\newblock \href {https://doi.org/10.1093/eurheartj/ehad191} {\path{doi:10.1093/eurheartj/ehad191}}.

\bibitem{Kashou2023}
A.~H. Kashou, P.~A. Noseworthy, T.~J. Beckman~et al., Impact of computer-interpreted {ECG}s on the accuracy of healthcare professionals, Current Problems in Cardiology 48~(11) (2023) 101989.
\newblock \href {https://doi.org/10.1016/j.cpcardiol.2023.101989} {\path{doi:10.1016/j.cpcardiol.2023.101989}}.

\bibitem{Ose2024}
B.~Ose, Z.~Sattar, A.~Gupta, C.~Toquica, C.~Harvey, A.~Noheria, Artificial intelligence interpretation of the electrocardiogram: A state-of-the-art review, Current Cardiology Reports 26~(6) (2024) 561–580.
\newblock \href {https://doi.org/10.1007/s11886-024-02062-1} {\path{doi:10.1007/s11886-024-02062-1}}.

\bibitem{Niemczyk2024}
S.~Niemczyk, J.~Fiegler-Rudol, M.~Migas, K.~Wągrowska, D.~Hochuł, J.~Talaska, T.~Klimczak, M.~Netkowska, Artificial intelligence in {ECG} analysis - future or present?, Emergency Medical Service 11~(2) (2024) 105–109.
\newblock \href {https://doi.org/10.36740/emems202402106} {\path{doi:10.36740/emems202402106}}.

\bibitem{Holmstrom2024}
L.~Holmstrom, H.~Chugh, K.~Nakamura, Z.~Bhanji, M.~Seifer, A.~Uy-Evanado, K.~Reinier, D.~Ouyang, S.~S. Chugh, An {ECG}-based artificial intelligence model for assessment of sudden cardiac death risk, Communications Medicine 4~(1) (Feb. 2024).
\newblock \href {https://doi.org/10.1038/s43856-024-00451-9} {\path{doi:10.1038/s43856-024-00451-9}}.

\bibitem{Kolk2024}
M.~Z. Kolk, S.~Ruipérez-Campillo, A.~A. Wilde, R.~E. Knops, S.~M. Narayan, F.~V. Tjong, Prediction of sudden cardiac death using artificial intelligence: Current status and future directions, Heart Rhythm (Sep. 2024).
\newblock \href {https://doi.org/10.1016/j.hrthm.2024.09.003} {\path{doi:10.1016/j.hrthm.2024.09.003}}.

\bibitem{Bacharova2023}
L.~Bacharova, P.~Chevalier, B.~Gorenek, C.~Jons, Y.~Li, E.~T. Locati, M.~Maanja, A.~R. Pérez‐Riera, P.~G. Platonov, A.~L.~P. Ribeiro, D.~Schocken, E.~Z. Soliman, J.~Svehlikova, L.~G. Tereshchenko, M.~Ugander, N.~Varma, Z.~Elena, T.~Ikeda, Ise/ishne expert consensus statement on the ecg diagnosis of left ventricular hypertrophy: The change of the paradigm, Annals of Noninvasive Electrocardiology 29~(1) (Nov. 2023).

\bibitem{Maron2018}
B.~J. Maron, E.~J. Rowin, M.~S. Maron, Global burden of hypertrophic cardiomyopathy, JACC: Heart Failure 6~(5) (2018) 376–378.
\newblock \href {https://doi.org/10.1016/j.jchf.2018.03.004} {\path{doi:10.1016/j.jchf.2018.03.004}}.

\bibitem{Martinez2024}
M.~Martinez, \href{https://ai.nejm.org/doi/full/10.1056/AI-S2400629}{Revolutionizing cardiology: {AI} in {ECG} analysis paves the way for better disease detection and treatment}, NEJM AI (2024).
\newline\urlprefix\url{https://ai.nejm.org/doi/full/10.1056/AI-S2400629}

\bibitem{Cheng2012}
Z.~Cheng, K.~Zhu, Z.~Tian, D.~Zhao, Q.~Cui, Q.~Fang, The findings of electrocardiography in patients with cardiac amyloidosis, Annals of Noninvasive Electrocardiology 18~(2) (2012) 157–162.
\newblock \href {https://doi.org/10.1111/anec.12018} {\path{doi:10.1111/anec.12018}}.

\bibitem{Collet2021}
J.~P. Collet, H.~Thiele, E.~Barbato, J.~Bauersachs, P.~Dendale, T.~Edvardsen, C.~P. Gale, A.~Jobs, E.~Lambrinou, J.~Mehilli, B.~Merkely, M.~Roffi, D.~Sibbing, A.~Kastrati, M.~A. Mamas, V.~Aboyans, D.~J. Angiolillo, H.~Bueno, R.~Bugiardini, R.~A. Byrne, S.~Castelletti, A.~Chieffo, V.~Cornelissen, F.~Crea, V.~Delgado, H.~Drexel, M.~Gierlotka, S.~Halvorsen, K.~H. Haugaa, E.~A. Jankowska, H.~A. Katus, T.~Kinnaird, J.~Kluin, V.~Kunadian, U.~Landmesser, C.~Leclercq, M.~Lettino, L.~Meinila, D.~Mylotte, G.~Ndrepepa, E.~Omerovic, R.~F. Pedretti, S.~E. Petersen, A.~S. Petronio, G.~Pontone, B.~A. Popescu, T.~Potpara, K.~K. Ray, F.~Luciano, D.~J. Richter, E.~Shlyakhto, I.~A. Simpson, M.~Sousa-Uva, R.~F. Storey, R.~M. Touyz, M.~Valgimigli, P.~Vranckx, R.~W. Yeh, O.~Barthélémy, D.~L. Bhatt, M.~Dorobantu, T.~Folliguet, M.~Gilard, P.~Jüni, B.~S. Lewis, E.~Meliga, C.~Mueller, F.~H. Rutten, G.~C. Siontis, 2020 {ESC} guidelines for the management of acute coronary syndromes in patients presenting without persistent {ST}-segment elevation (2021).
\newblock \href {https://doi.org/10.1093/eurheartj/ehaa575} {\path{doi:10.1093/eurheartj/ehaa575}}.

\bibitem{pmid29084731}
S.~M. Al-Khatib, W.~G. Stevenson, M.~J. Ackerman, W.~J. Bryant, D.~J. Callans, A.~B. Curtis, B.~J. Deal, T.~Dickfeld, M.~E. Field, G.~C. Fonarow, A.~M. Gillis, M.~A. Hlatky, C.~B. Granger, S.~C. Hammill, J.~A. Joglar, G.~N. Kay, D.~D. Matlock, R.~J. Myerburg, R.~L. Page, {2017 {A}{H}{A}/{A}{C}{C}/{H}{R}{S} {G}uideline for {M}anagement of {P}atients {W}ith {V}entricular {A}rrhythmias and the {P}revention of {S}udden {C}ardiac {D}eath: {A} {R}eport of the {A}merican {C}ollege of {C}ardiology/{A}merican {H}eart {A}ssociation {T}ask {F}orce on {C}linical {P}ractice {G}uidelines and the {H}eart {R}hythm {S}ociety}, Circulation (Oct 2017).

\bibitem{pmid27029760}
S.~G. Priori, C.~Blomstrom-Lundqvist, A.~Mazzanti, N.~Blom, M.~Borggrefe, J.~Camm, P.~M. Elliott, D.~Fitzsimons, R.~Hatala, G.~Hindricks, P.~Kirchhof, K.~Kjeldsen, K.~H. Kuck, A.~Hernandez-Madrid, N.~Nikolaou, T.~M. Norekval, C.~Spaulding, D.~J. Van~Veldhuisen, {[2015 {E}{S}{C} {G}uidelines for the management of patients with ventricular arrhythmias and the prevention of sudden cardiac death.]}, G Ital Cardiol (Rome) 17~(2) (2016) 108--170.

\bibitem{Anderson2019}
R.~D. Anderson, S.~Kumar, R.~Parameswaran, G.~Wong, A.~Voskoboinik, H.~Sugumar, T.~Watts, P.~B. Sparks, J.~B. Morton, A.~McLellan, P.~M. Kistler, J.~Kalman, G.~Lee, Differentiating right-and left-sided outflow tract ventricular arrhythmias: Classical {ECG} signatures and prediction algorithms, Circulation 12 (2019) e007392.

\bibitem{Frisk2024}
C.~Frisk, S.~Das, M.~J. Eriksson, A.~Walentinsson, M.~Corbascio, C.~Hage, C.~Kumar, M.~Ekström, E.~Maret, H.~Persson, C.~Linde, B.~Persson, Cardiac biopsies reveal differences in transcriptomics between left and right ventricle in patients with or without diagnostic signs of heart failure, Scientific Reports 14 (2024) 5811.
\newblock \href {https://doi.org/10.1038/s41598-024-56025-1} {\path{doi:10.1038/s41598-024-56025-1}}.

\bibitem{Sedova2023}
K.~A. Sedova, P.~M. van Dam, M.~Blahova, L.~Necasova, J.~Kautzner, Localization of the ventricular pacing site from {BSPM} and standard 12-lead {ECG}: a comparison study, Scientific Reports 13~(1) (2023).

\bibitem{Cluitmans2018}
M.~Cluitmans, D.~H. Brooks, R.~MacLeod, O.~D\"{o}ssel, M.~S. Guillem, P.~M. van Dam, J.~Svehlikova, B.~He, J.~Sapp, L.~Wang, L.~Bear, Validation and opportunities of electrocardiographic imaging: From technical achievements to clinical applications, Frontiers in Physiology 9 (Sep. 2018).
\newblock \href {https://doi.org/10.3389/fphys.2018.01305} {\path{doi:10.3389/fphys.2018.01305}}.

\bibitem{Li2024}
L.~Li, J.~Camps, B.~Rodriguez, V.~Grau, Solving the inverse problem of electrocardiography for cardiac digital twins: A survey (2024).
\newblock \href {https://doi.org/10.48550/ARXIV.2406.11445} {\path{doi:10.48550/ARXIV.2406.11445}}.

\bibitem{Bear2018}
L.~R. Bear, I.~J. LeGrice, G.~B. Sands, N.~A. Lever, D.~S. Loiselle, D.~J. Paterson, L.~K. Cheng, B.~H. Smaill, How accurate is inverse electrocardiographic mapping?: A systematic in vivo evaluation, Circulation: Arrhythmia and Electrophysiology 11~(5) (May 2018).
\newblock \href {https://doi.org/10.1161/circep.117.006108} {\path{doi:10.1161/circep.117.006108}}.

\bibitem{Potyagaylo2019}
D.~Potyagaylo, M.~Chmelevsky, P.~V. Dam, M.~Budanova, S.~Zubarev, T.~Treshkur, D.~Lebedev, {ECG} adapted fastest route algorithm to localize the ectopic excitation origin in {CRT} patients, Frontiers in Physiology 10 (2019).
\newblock \href {https://doi.org/10.3389/fphys.2019.00183} {\path{doi:10.3389/fphys.2019.00183}}.

\bibitem{BibbinsDomingo}
K.~Bibbins-Domingo, A.~Helman, Improving Representation in Clinical Trials and Research: Building Research Equity for Women and Underrepresented Groups, 2022.
\newblock \href {https://doi.org/10.17226/26479} {\path{doi:10.17226/26479}}.

\bibitem{Buchner2019}
T.~Buchner, On the physical nature of biopotentials, their propagation and measurement, Physica A: Statistical Mechanics and its Applications 525 (2019) 85–95.
\newblock \href {https://doi.org/10.1016/j.physa.2019.03.056} {\path{doi:10.1016/j.physa.2019.03.056}}.

\bibitem{Buchner2023}
T.~Buchner, M.~Zajdel, K.~Peczalski, P.~Nowak, Finite velocity of {ECG} signal propagation: preliminary theory, results of a pilot experiment and consequences for medical diagnosis, Scientific Reports 13~(1) (Mar. 2023).
\newblock \href {https://doi.org/10.1038/s41598-023-29904-2} {\path{doi:10.1038/s41598-023-29904-2}}.

\bibitem{Peczalski2024}
K.~Peczalski, J.~Sobiech, T.~Buchner, T.~Kornack, E.~Foley, D.~Janczak, M.~Jakubowska, D.~Newby, N.~Ford, M.~Zajdel, Synchronous recording of magnetocardiographic and electrocardiographic signals, Scientific Reports 14~(1) (Feb. 2024).
\newblock \href {https://doi.org/10.1038/s41598-024-54126-5} {\path{doi:10.1038/s41598-024-54126-5}}.

\bibitem{Scharf}
G.~Scharf, L.~Dang, C.~Scharf, Electrophysiology of living organs from first principles. (2010).
\newblock \href {http://arxiv.org/abs/arXiv:bio-ph/1006.3453} {\path{arXiv:arXiv:bio-ph/1006.3453}}, \href {https://doi.org/10.1023/A:1004600603161} {\path{doi:10.1023/A:1004600603161}}.

\bibitem{bembook}
J.~Malmivuo, R.~Plonsey, Bioelectromagnetism--Principles and applications of bioelectric and biomagnetic fields, Oxford University Press, New York, 1995, 1995.

\bibitem{Hossenfelder2019}
S.~Hossenfelder, Lost in math, Basic Civitas Books, New York, NY, 2019.

\bibitem{pmid26082525}
Y.~Rudy, {{T}he forward problem of electrocardiography revisited}, Circ Arrhythm Electrophysiol 8~(3) (2015) 526--528.

\bibitem{Cluitmans}
M.~Cluitmans, Noninvasive reconstruction of cardiac electrical activity: Mathematical innovation, in vivo validation and human application, Ph.D. thesis (2016).

\bibitem{Franz1999}
M.~Franz, Current status of monophasic action potential recording: theories, measurements and interpretations, Cardiovascular Research 41~(1) (1999) 25–40.

\bibitem{Barr1977}
R.~C. Barr, M.~Ramsey, M.~S. Spach, {Relating epicardial to body surface potential distributions by means of transfer coefficients based on geometry measurements}, IEEE Transactions on biomedical engineering~(1) (1977) 1--11.

\bibitem{BAMS2024}
J.~Strzałkowski, P.~Polak, T.~Buchner, Capacitive coupling between the heart and tissue and its mathematical representation in future problems, Bio-Algorithms and Med-Systems 20~(1) (2024) 159–168.
\newblock \href {https://doi.org/10.5604/01.3001.0054.9268} {\path{doi:10.5604/01.3001.0054.9268}}.

\bibitem{Durrer}
D.~Durrer, R.~T. van Dam, G.~E. Freud, M.~J. Janse, F.~L. Meijler, R.~C. Arzbaecher, {{T}otal excitation of the isolated human heart}, Circulation 41~(6) (1970) 899--912.

\bibitem{Orini2016}
M.~Orini, P.~Taggart, N.~Srinivasan, M.~Hayward, P.~D. Lambiase, Interactions between activation and repolarization restitution properties in the intact human heart: In-vivo whole-heart data and mathematical description, PLoS One 11~(9) (2016) e0161765.
\newblock \href {https://doi.org/10.1371/journal.pone.0161765} {\path{doi:10.1371/journal.pone.0161765}}.

\bibitem{Fast1997}
V.~G. Fast, A.~G. Kléber, Role of wavefront curvature in propagation of cardiac impulse, Cardiovascular Research 33~(2) (1997) 258–271.
\newblock \href {https://doi.org/10.1016/s0008-6363(96)00216-7} {\path{doi:10.1016/s0008-6363(96)00216-7}}.

\bibitem{Hanson2009}
B.~Hanson, P.~Sutton, N.~Elameri, M.~Gray, H.~Critchley, J.~S. Gill, P.~Taggart, Interaction of activation–repolarization coupling and restitution properties in humans, Circulation: Arrhythmia and Electrophysiology 2~(2) (2009) 162–170.
\newblock \href {https://doi.org/10.1161/circep.108.785352} {\path{doi:10.1161/circep.108.785352}}.

\bibitem{Qu2000}
Z.~Qu, F.~Xie, A.~Garfinkel, J.~N. Weiss, Origins of spiral wave meander and breakup in a two-dimensional cardiac tissue model, Annals of Biomedical Engineering 28~(7) (2000) 755–771.
\newblock \href {https://doi.org/10.1114/1.1289474} {\path{doi:10.1114/1.1289474}}.

\bibitem{Weiss1999}
J.~N. Weiss, A.~Garfinkel, H.~S. Karagueuzian, Z.~Qu, P.-S. Chen, Chaos and the transition to ventricular fibrillation: A new approach to antiarrhythmic drug evaluation, Circulation 99~(21) (1999) 2819–2826.
\newblock \href {https://doi.org/10.1161/01.cir.99.21.2819} {\path{doi:10.1161/01.cir.99.21.2819}}.

\bibitem{Jabr2016}
R.~I. Jabr, F.~S. Hatch, S.~C. Salvage, A.~Orlowski, P.~D. Lampe, C.~H. Fry, Regulation of gap junction conductance by calcineurin through {C}x43 phosphorylation: implications for action potential conduction, Pfl\"{u}gers Archiv - European Journal of Physiology 468~(11–12) (2016) 1945–1955.
\newblock \href {https://doi.org/10.1007/s00424-016-1885-7} {\path{doi:10.1007/s00424-016-1885-7}}.

\bibitem{Cuellar}
A.~A. Cuellar, C.~M. Lloyd, P.~F. Nielsen, D.~P. Bullivant, D.~P. Nickerson, P.~J. Hunter, An overview of {CellML} 1.1, a biological model description language, Simulation 79~(12) (2003) 740--747.
\newblock \href {https://doi.org/10.1177/0037549703040939} {\path{doi:10.1177/0037549703040939}}.

\bibitem{Ten_Tusscher2004-ns}
K.~H. W.~J. ten Tusscher, D.~Noble, P.~J. Noble, A.~V. Panfilov, A model for human ventricular tissue, Am. J. Physiol. Heart Circ. Physiol. 286~(4) (2004) H1573--89.

\bibitem{LuoRudy}
C.~H. Luo, Y.~Rudy, {{A} dynamic model of the cardiac ventricular action potential. {I}. {S}imulations of ionic currents and concentration changes}, Circ. Res. 74~(6) (1994) 1071--1096.

\bibitem{Clerx2016}
M.~Clerx, P.~Collins, E.~de~Lange, P.~G. Volders, Myokit: A simple interface to cardiac cellular electrophysiology, Progress in Biophysics and Molecular Biology 120~(1–3) (2016) 100–114.
\newblock \href {https://doi.org/10.1016/j.pbiomolbio.2015.12.008} {\path{doi:10.1016/j.pbiomolbio.2015.12.008}}.

\bibitem{Gargiulo}
G.~D. Gargiulo, A.~L. McEwan, P.~Bifulco, M.~Cesarelli, C.~Jin, J.~Tapson, A.~Thiagalingam, A.~van Schaik, Towards true unipolar {ECG} recording without the {W}ilson central terminal (preliminary results), Physiological Measurement 34~(9) (2013) 991.

\bibitem{Kappadan2023}
V.~Kappadan, A.~Sohi, U.~Parlitz, S.~Luther, I.~Uzelac, F.~Fenton, N.~S. Peters, J.~Christoph, F.~S. Ng, Optical mapping of contracting hearts, The Journal of Physiology 601~(8) (2023) 1353–1370.
\newblock \href {https://doi.org/10.1113/jp283683} {\path{doi:10.1113/jp283683}}.

\bibitem{OKAMOTO1998291}
Y.~Okamoto, S.~Mashima, The zero potential and {W}ilson's central terminal in electrocardiography, Bioelectrochemistry and Bioenergetics 47~(2) (1998) 291 -- 295.
\newblock \href {https://doi.org/https://doi.org/10.1016/S0302-4598(98)00201-3} {\path{doi:https://doi.org/10.1016/S0302-4598(98)00201-3}}.

\bibitem{WILSON1934447}
F.~N. Wilson, F.~D. Johnston, A.~Macleod, P.~S. Barker, Electrocardiograms that represent the potential variations of a single electrode, American Heart Journal 9~(4) (1934) 447 -- 458.
\newblock \href {https://doi.org/https://doi.org/10.1016/S0002-8703(34)90093-4} {\path{doi:https://doi.org/10.1016/S0002-8703(34)90093-4}}.

\bibitem{pytorch}
J.~Ansel, E.~Yang, H.~He, N.~Gimelshein~et al., Pytorch 2: Faster machine learning through dynamic {P}ython bytecode transformation and graph compilation, in: Proceedings of the 29th ACM International Conference on Architectural Support for Programming Languages and Operating Systems, Volume 2, ASPLOS '24, Association for Computing Machinery, New York, NY, USA, 2024, p. 929–947.
\newblock \href {https://doi.org/10.1145/3620665.3640366} {\path{doi:10.1145/3620665.3640366}}.

\bibitem{ptbxl}
P.~Wagner, N.~Strodthoff, R.-D. Bousseljot, W.~Samek, T.~Schaeffter, {PTB-XL}, a large publicly available electrocardiography dataset (2022).
\newblock \href {https://doi.org/10.13026/KFZX-AW45} {\path{doi:10.13026/KFZX-AW45}}.

\bibitem{neurokit}
D.~Makowski, T.~Pham, Z.~J. Lau, J.~C. Brammer, F.~Lespinasse, H.~Pham, C.~Sch\"{o}lzel, S.~H.~A. Chen, {NeuroKit2}: {A} {P}ython toolbox for neurophysiological signal processing, Behavior Research Methods 53~(4) (2021) 1689–1696.
\newblock \href {https://doi.org/10.3758/s13428-020-01516-y} {\path{doi:10.3758/s13428-020-01516-y}}.

\bibitem{ptb}
R.-D. Bousseljot, D.~Kreiseler, A.~Schnabel, The {PTB} diagnostic {ECG} database, Biomedizinische Technik / Biomedical Engineering 40~(1) (1995) 317--318.

\bibitem{2020SciPy}
P.~Virtanen, R.~Gommers, T.~E. Oliphant, M.~Haberland, T.~Reddy, D.~Cournapeau, E.~Burovski, P.~Peterson, W.~Weckesser~et al., {{SciPy} 1.0: Fundamental Algorithms for Scientific Computing in {P}ython}, Nature Methods 17 (2020) 261--272.
\newblock \href {https://doi.org/10.1038/s41592-019-0686-2} {\path{doi:10.1038/s41592-019-0686-2}}.

\bibitem{PARDEE}
H.~Pardee, An electrocardiographic sign of coronary artery obstruction, Archives of Internal Medicine 26~(2) (1920) 244--257.
\newblock \href {https://doi.org/10.1001/archinte.1920.00100020113007} {\path{doi:10.1001/archinte.1920.00100020113007}}.

\bibitem{Waddingham2022}
P.~H. Waddingham, J.~O. Mangual, M.~Orini, N.~Badie, A.~Muthumala, S.~Sporton, L.~C. McSpadden, P.~D. Lambiase, A.~W.~C. Chow, Electrocardiographic imaging demonstrates electrical synchrony improvement by dynamic atrioventricular delays in patients with left bundle branch block and preserved atrioventricular conduction, EP Europace 25~(2) (2022) 536–545.
\newblock \href {https://doi.org/10.1093/europace/euac224} {\path{doi:10.1093/europace/euac224}}.

\bibitem{Elliott2021}
M.~K. Elliott, J.~Blauer, V.~S. Mehta, B.~S. Sidhu, J.~Gould, T.~Jackson, B.~Sieniewicz, S.~Niederer, S.~Ghosh, C.~A. Rinaldi, Comparison of electrical dyssynchrony parameters between electrocardiographic imaging and a simulated {ECG} belt, Journal of Electrocardiology 68 (2021) 117–123.
\newblock \href {https://doi.org/10.1016/j.jelectrocard.2021.08.003} {\path{doi:10.1016/j.jelectrocard.2021.08.003}}.

\bibitem{Verstappen2024}
A.~A. Verstappen, R.~Hautvast, P.~Jurak, F.~A. Bracke, L.~M. Rademakers, Ventricular dyssynchrony imaging, echocardiographic and clinical outcomes of left bundle branch pacing and biventricular pacing, Indian Pacing and Electrophysiology Journal 24~(3) (2024) 140–146.
\newblock \href {https://doi.org/10.1016/j.ipej.2024.04.007} {\path{doi:10.1016/j.ipej.2024.04.007}}.

\bibitem{Nguyn2024}
U.~C. Nguy\^en, J.~H.~J. Rijks, F.~Plesinger, L.~M. Rademakers, J.~Luermans, K.~C. Smits, A.~M.~W. van Stipdonk, F.~W. Prinzen, K.~Vernooy, J.~Halamek, K.~Curila, P.~Jurak, Ultra-high-frequency {ECG} in cardiac pacing and cardiac resynchronization therapy: From technical concept to clinical application, Journal of Cardiovascular Development and Disease 11~(3) (2024) 76.
\newblock \href {https://doi.org/10.3390/jcdd11030076} {\path{doi:10.3390/jcdd11030076}}.

\bibitem{Jurak2017}
P.~Jurak, J.~Halamek, J.~Meluzin, F.~Plesinger, T.~Postranecka, J.~Lipoldova, M.~Novak, V.~Vondra, I.~Viscor, L.~Soukup, P.~Klimeš, P.~Vesely, J.~Sumbera, K.~Zeman, R.~Asirvatham, J.~Tri, S.~Asirvatham, P.~Leinveber, Ventricular dyssynchrony assessment using ultra-high frequency {ECG} technique, Journal of Interventional Cardiac Electrophysiology: an International Journal of Arrhythmias and Pacing 49 (09 2017).
\newblock \href {https://doi.org/10.1007/s10840-017-0268-0} {\path{doi:10.1007/s10840-017-0268-0}}.

\bibitem{Isaksen2021}
J.~L. Isaksen, J.~Ghouse, C.~Graff, M.~S. Olesen, A.~G. Holst, A.~Pietersen, J.~B. Nielsen, M.~W. Skov, J.~K. Kanters, Electrocardiographic {T}-wave morphology and risk of mortality, International Journal of Cardiology 328 (2021) 199–205.
\newblock \href {https://doi.org/10.1016/j.ijcard.2020.12.016} {\path{doi:10.1016/j.ijcard.2020.12.016}}.

\bibitem{KrijgerJurez2023}
C.~Krijger~Ju{\'{a}}rez, A.~S. Amin, J.~A. Offerhaus, C.~R. Bezzina, B.~J. Boukens, Cardiac repolarization in health and disease, JACC: Clinical Electrophysiology 9~(1) (2023) 124–138.
\newblock \href {https://doi.org/10.1016/j.jacep.2022.09.017} {\path{doi:10.1016/j.jacep.2022.09.017}}.

\bibitem{Xue2010}
J.~Xue, Y.~Chen, X.~Han, W.~Gao, Electrocardiographic morphology changes with different type of repolarization dispersions, Journal of Electrocardiology 43~(6) (2010) 553–559.
\newblock \href {https://doi.org/10.1016/j.jelectrocard.2010.07.011} {\path{doi:10.1016/j.jelectrocard.2010.07.011}}.

\bibitem{Ikeda2021}
T.~Ikeda, Right bundle branch block: Current considerations, Current Cardiology Reviews 17~(1) (2021) 24–30.
\newblock \href {https://doi.org/10.2174/1573403x16666200708111553} {\path{doi:10.2174/1573403x16666200708111553}}.

\bibitem{Calle2022}
S.~Calle, F.~Timmermans, J.~De~Pooter, Defining left bundle branch block according to the new 2021 {E}uropean {S}ociety of {C}ardiology criteria, Netherlands Heart Journal 30~(11) (2022) 495–498.
\newblock \href {https://doi.org/10.1007/s12471-022-01697-5} {\path{doi:10.1007/s12471-022-01697-5}}.

\bibitem{Kloosterman2024}
M.~Kloosterman, K.~P. Loh, T.~A.~B. van Veen, Left bundle branch block-induced cardiomyopathy: A distinctive form of cardiomyopathy that might require a dedicated form of treatment, Heart Rhythm 21~(8) (2024) 1380--1381.

\bibitem{Pipberger1961-fi}
H.~V. Pipberger, S.~M. Bialek, J.~K. Perloff, H.~W. Schnaper, Correlation of clinical information in the standard 12-lead {ECG} and in a corrected orthogonal 3-lead {ECG}, Am. Heart J. 61~(1) (1961) 34--43.

\bibitem{pmid668061}
W.~T. Miller, D.~B. Geselowitz, {{S}imulation studies of the electrocardiogram. {I}. {T}he normal heart}, Circ. Res. 43~(2) (1978) 301--315.

\bibitem{Kaufman}
W.~Kaufman, F.~D. Johnston, The electrical conductivity of the tissues near the heart and its bearing on the distribution of the cardiac action currents, American Heart Journal 26~(1) (1943) 42--54.
\newblock \href {https://doi.org/10.1016/S0002-8703(43)90050-X} {\path{doi:10.1016/S0002-8703(43)90050-X}}.

\bibitem{Wilson}
F.~N. Wilson, A.~G. Macleod, P.~S. Barker, {The distribution of the action currents produced by heart muscle and other excitable tissues immersed in extensive conducting media}, J. Gen. Physiol. 16~(3) (1933) 423--456.

\bibitem{FRANK}
E.~Frank, {{G}eneral theory of heart-vector projection}, Circ. Res. 2~(3) (1954) 258--270.

\bibitem{Roth}
B.~J. Roth, The electrical conductivity of tissues, CRC Press LLC, Boca Raton, 2000.

\bibitem{Brody}
D.~A. Brody, {{A} theoretical analysis of intracavitary blood mass influence on the heart-lead relationship}, Circ. Res. 4~(6) (1956) 731--738.

\bibitem{pmid6048873}
A.~C. Barnard, I.~M. Duck, M.~S. Lynn, {{T}he application of electromagnetic theory to electrocardiology. {I}. {D}erivation of the integral equations}, Biophys. J. 7~(5) (1967) 443--462.

\bibitem{Oosterom}
A.~van Oosterom, {{G}enesis of the {T} wave as based on an equivalent surface source model}, J Electrocardiol 34 Suppl (2001) 217--227.

\bibitem{pmid19465555}
M.~Potse, A.~Vinet, T.~Opthof, R.~Coronel, {{V}alidation of a simple model for the morphology of the {T} wave in unipolar electrograms}, Am. J. Physiol. Heart Circ. Physiol. 297~(2) (2009) 792--801.

\bibitem{pmid22282106}
M.~Potse, {{M}athematical modeling and simulation of ventricular activation sequences: implications for cardiac resynchronization therapy}, J Cardiovasc Transl Res 5~(2) (2012) 146--158.

\bibitem{pmid22006012}
T.~F. Oostendorp, P.~F. van Dessel, R.~Coronel, C.~Belterman, A.~C. Linnenbank, I.~H. van Schie, A.~van Oosterom, P.~Oosterhoff, P.~M. van Dam, J.~M. de~Bakker, {{N}oninvasive detection of epicardial and endocardial activity of the heart}, Neth Heart J 19~(11) (2011) 488--491.

\bibitem{Geselowitz}
D.~Geselowitz, {{O}n the theory of the electrocardiogram}, Proc. IEEE 77~(6) (1989) 857--876.

\bibitem{pmid25896779}
M.~J. Cluitmans, R.~L. Peeters, R.~L. Westra, P.~G. Volders, {{N}oninvasive reconstruction of cardiac electrical activity: update on current methods, applications and challenges}, Neth Heart J 23~(6) (2015) 301--311.

\bibitem{HECKERT1961657}
E.~W. Heckert, W.~R. Cook, S.~Krause, The clinical value of vectorcardiography, The American Journal of Cardiology 7~(5) (1961) 657 -- 660.
\newblock \href {https://doi.org/https://doi.org/10.1016/0002-9149(61)90449-0} {\path{doi:https://doi.org/10.1016/0002-9149(61)90449-0}}.

\bibitem{Dehghani2010}
N.~Dehghani, C.~B{\'{e}}dard, S.~S. Cash, E.~Halgren, A.~Destexhe, Comparative power spectral analysis of simultaneous elecroencephalographic and magnetoencephalographic recordings in humans suggests non-resistive extracellular media, Journal of Computational Neuroscience 29~(3) (2010) 405--421.
\newblock \href {https://doi.org/10.1007/s10827-010-0263-2} {\path{doi:10.1007/s10827-010-0263-2}}.

\bibitem{Pereda2014}
A.~E. Pereda, Electrical synapses and their functional interactions with chemical synapses, Nature Reviews Neuroscience 15~(4) (2014) 250–263.
\newblock \href {https://doi.org/10.1038/nrn3708} {\path{doi:10.1038/nrn3708}}.

\bibitem{Pietak}
A.~Pietak, M.~Levin, {{E}xploring instructive physiological signaling with the bioelectric tissue simulation engine}, Front Bioeng Biotechnol 4 (2016) 55.

\bibitem{Halnes}
G.~Halnes, I.~Ostby, K.~H. Pettersen, S.~W. Omholt, G.~T. Einevoll, {{E}lectrodiffusive model for astrocytic and neuronal ion concentration dynamics}, PLoS Comput. Biol. 9~(12) (2013) e1003386.

\bibitem{katz}
L.~Katz, A.~Bohning, I.~Gutman, K.~Jochim, H.~Korey, P.~Ocko, M.~Robinow, {{C}oncerning a new concept of the genesis of the electrocardiogram}, Am. Heart J. 13~(1) (1937) 17--35.

\bibitem{merrill2005171}
D.~R. Merrill, M.~Bikson, J.~G. Jefferys, Electrical stimulation of excitable tissue: design of efficacious and safe protocols, Journal of Neuroscience Methods 141~(2) (2005) 171 -- 198.

\bibitem{Lim2014}
Y.~G. Lim, J.~S. Lee, S.~M. Lee, H.~J. Lee, K.~S. Park, Capacitive measurement of {ECG} for ubiquitous healthcare, Annals of Biomedical Engineering 42~(11) (2014) 2218–2227.
\newblock \href {https://doi.org/10.1007/s10439-014-1069-6} {\path{doi:10.1007/s10439-014-1069-6}}.

\bibitem{Khalili2024}
M.~Khalili, H.~GholamHosseini, A.~Lowe, M.~M.~Y. Kuo, Motion artifacts in capacitive {ECG} monitoring systems: a review of existing models and reduction techniques, Medical \& Biological Engineering \& Computing (Jul. 2024).
\newblock \href {https://doi.org/10.1007/s11517-024-03165-1} {\path{doi:10.1007/s11517-024-03165-1}}.

\bibitem{feynman1965flp}
R.~Feynman, R.~Leighton, M.~Sands, E.~Hafner, {The Feynman Lectures on Physics; Vol. I}, Vol.~33, AAPT, 1965.

\end{thebibliography}

\clearpage

\appendix
\section{Introduction into Molecular Theory of Biopotentials}\label{app:Molecular}

As the foundations of the PhysECG algorithm are grounded in the physics of the ECG, it seems helpful to extend the argumentation to justify its basic assumptions.

There have been numerous attempts to reliably model the cardiac activity, with several interesting ideas and interpretations of the obtained results \cite{pmid668061,Kaufman,Wilson,FRANK,Roth,Brody,pmid6048873,Oosterom,pmid19465555,pmid22282106,pmid22006012,Geselowitz,pmid26082525,pmid25896779,Cluitmans,Cluitmans2018,Li2024}. The resultant, widely accepted geometric theory of ECG, very roughly speaking, considers potentials in leads as projections of the heart's electrical vector \cite{HECKERT1961657} and the corresponding cardiac dipole onto Einthoven's triangle in the frontal plane or the spatial directions of precordial electrodes in the transverse plane \cite{Geselowitz,bembook}. 

The molecular theory for biopotentials (MTB), which constitutes the theoretical foundation of the PhysECG algorithm, refers to the physical nature of the ECG, which is one level below the purely mathematical description of global quantities, such as potential described by monodomain equations and Laplace equation. This line of modeling led to many unprecedented successes - c.f. \cite{Cluitmans2018,Li2024} for a review, which we will not undermine. Here, we would ask: What can be added value of widening the modeling scope, and what is the added value for ECGI? Finally, we would like to identify medical and technical cases where the MTB offers the simplest possible explanation.

From a physical point of view, it is a truism that each spatiotemporal electrostatic (quasistatic) potential has an equivalent distribution of charge $\rho$. Free charges at fixed positions are treated as sources, whereas the charges, which can be repositioned as a result of the action of the sources, are referred to as bound charges. Their alternative name is induced charges, which underline their role in the electrostatic induction process, described mathematically by displacement vector $\vec{D}$. A measurable manifestation of this vector is, by definition, the nonzero density of induced surface charge $\sigma$. The presence of $\sigma \not= 0$ results from the action of an external electric field on all objects that are dielectric or have a dielectric component of electric response. A living body is no exception to this rule, as the capacitive response was quite recently observed \cite{Dehghani2010}. Also, it is typical that charges in nature are paired, and their sum over a certain space (domain) $\Omega$ is typically zero. This does not mean the spatial distribution of charge $\rho$ in this space satisfies: $\rho \equiv 0|_{\Omega}$ - c.f. \cite{Buchner2019} for further discussion. 

The potential of induced surface charge is the entity directly measured by the ECG electrode, which gets coupled by charge transfer processes to the electric field that controls the gate of the FET transistor in the operational amplifier.
We do not consider any layers deeper than the epicardium directly, as the process of coupling through the pericardial sac favors proximal layers. Instead, we interpret the epicardial potentials as the end result of their intrinsic activity and the effect of deeper layers, expressed mainly through passive physical coupling through the electrically polarized cells and passive coupling through the altered levels of electrolytes in the extracellular space. The motivation here is the concept of electrical synapse \cite{Pereda2014} and the electrolyte concentration-based approach to cellular modeling \cite{Pietak,Halnes}. The state of the electrode , in turn, is directly affected only by the area below the electrode, to which the signal is delivered using physical transport mechanisms. This opinion is justified by such techniques as unipolar potentials \cite{Franz1999}, where the alterations of the tissue below the electrode, induced by suction and local ischemia, were shown to completely abolish the observability of cardiac potentials, while the spatial position of the electrode was not altered. The severing of the connective tissue in the vicinity of the dog's heart reduced the ECG in limb leads, although the heart was still in anatomical position \cite{katz}.
Here, we do not follow the approach by Scharf et al., as we do not consider current dipoles as sources \cite{Scharf}. However, we do appreciate the introduction of charge density and volume polarization vector.

The signal propagates through capacitive coupling (on non-conductive barriers) and tissue polarization to the skin surface, provided that the dielectric constants in segments permit this \cite{Buchner2019}. Finally, the skin surface polarizes the electrode, incorporating both Faradaic (redox reactions) and non-Faradaic (double layer) mechanisms \cite{merrill2005171}. The capacitive measurements of the ECG, which recently became very popular, show that galvanic contact is not a prerequisite for the observability of the ECG. They also deliver an indirect clue that the normal component of the electric field (a.k.a. potential gradient) on the body surface is nonzero, as it passes through the dielectric layer to polarize the electrode \cite{Lim2014,Khalili2024}. 

The presence of a charge in a passive medium is neglected in all approaches that refer to the Laplace equation. In our opinion, it is beneficial to take one step back and assume that the total charge is indeed zero, but the spatiotemporal distribution of ions and coions is important, constituting the most important mechanism of charge transfer. For instance, electrode models that assume AC coupling through a capacitor, e.g., \cite{Khalili2024}, require charges at the input to the circuit; this cannot be modeled using Laplace-based models. The main component of the tissue response is the volume ionic current confined to multiple finite spaces: the cells, and other mesoscopic structures, which justifies the idea of a volume conductor.

The induction of an inductive charge results from tissue polarization processes aimed at minimizing the tissue energy in the external electric field (relaxation of energy received from the source). Therefore, the tissue response is such that the relaxation process occurs as quickly as possible, i.e., when the signal reaches the skin in minimum time - c.f the Feynman explanation of Snell law \cite[ch.~31]{feynman1965flp} - note, that the tissue passage time is non-negligible \cite{Buchner2023}. Consequently, for precordial electrodes, we observe signal differences between adjacent channels despite the physical proximity of the electrodes. Thus, for the precordial leads, we have assumed that the signal can be considered a local representation of the epicardium's state rather than a weighted spatial average of a whole heart.

In the case of limb electrodes, it is impossible to identify the epicardial segment coupled to them directly; they represent the broad result of many cardiac segments.
\end{document}